\documentclass[graybox, secnum]{svmult}

\usepackage{mathptmx}       
\usepackage{helvet}         
\usepackage{courier}        
\usepackage{type1cm}        
\usepackage{makeidx}         
\usepackage{graphicx}        
\usepackage{multicol}        
\usepackage[bottom]{footmisc}
\usepackage{hyperref}        
\usepackage{soul}            
\usepackage{xcolor}
\definecolor{bluecite}{HTML}{0875b7}

\hypersetup{
	colorlinks=true,
	linkcolor=bluecite,
	urlcolor=bluecite,
	citecolor=bluecite
}
\usepackage[square,numbers,compress]{natbib}
\makeindex             

\usepackage{amsmath,amsfonts,amssymb}
\usepackage{subcaption}
\usepackage{bbold}
\usepackage{tensor,slashed}
\usepackage[T1]{fontenc}
\usepackage[utf8]{inputenc}

\newcommand{\eg}{{\textit{e.g.}}}
\newcommand{\ie}{{\textit{i.e.}}}
\newcommand{\dd}{\ensuremath{\mathrm{d}}}
\newcommand{\dalembertian}{\ensuremath{\square}}
\newcommand{\CD}{\ensuremath{\nabla}}
\renewcommand{\imath}{\ensuremath{\mathrm{i}}}

\newcommand{\GN}{\ensuremath{G_N}}
\newcommand{\p}{\ensuremath{\partial}}
\newcommand{\mans}{\ensuremath{s}}
\newcommand{\mant}{\ensuremath{t}}
\newcommand{\manu}{\ensuremath{u}}

\newcommand{\cO}{\mathcal{O}}

\usepackage{ulem}

\begin{document}
	\title*{Form Factors \\ in Asymptotically Safe Quantum Gravity}
	\author{Benjamin Knorr, Chris Ripken, and Frank Saueressig\thanks{corresponding author}}
	\institute{Benjamin Knorr \at Perimeter Institute for Theoretical Physics, 31 Caroline Street North, Waterloo, ON N2L 2Y5, Canada, and \\
	Nordita, Stockholm University and KTH Royal Institute of Technology, Hannes Alfvéns väg 12, SE-106 91 Stockholm, Sweden \\ \email{benjamin.knorr@su.se}
		\and Chris Ripken \at Institute of Physics (THEP), University of Mainz, Staudingerweg 7, 55128 Mainz, Germany \\ \email{aripken@uni-mainz.de} \and 
		Frank Saueressig \at Institute for Mathematics, Astrophysics and Particle Physics (IMAPP), Radboud University Nijmegen, Heyendaalseweg 135, 6525 AJ Nijmegen, The Netherlands \\ \email{f.saueressig@science.ru.nl}}
	%
	%
	\maketitle
	\abstract{Asymptotic Safety implies that observables including scattering amplitudes remain finite at the highest energy scales. Traditionally, this feature is connected to an interacting fixed point of the Wilsonian renormalization group that provides the high-energy completion of the theory. In this article, we discuss a different facet of Asymptotic Safety, reviewing its imprint on the quantum effective action. We start with a concise introduction to the form factor framework for gravity and gravity-matter systems, before giving an encyclopaedic overview of gravity-mediated scattering amplitudes derived from the quantum effective action. We illustrate the use of the framework based on form factors appearing in the context of quadratic gravity and Asymptotic Safety, before making the connections to positivity bounds derived for low-energy effective field theories and the computation of form factors from first principles. We conclude that the form factor framework offers a unique, unifying perspective on quantum gravity. In particular, it plays a pivotal role in determining the phenomenological consequences of Asymptotic Safety at the level of observables.}
	
	\section*{Keywords}
	Asymptotic Safety, quantum effective action, scattering amplitudes, form factors, functional renormalization group, graviton propagator, gravity-matter couplings
	\section{Introduction}
	\label{sec.intro}
	The gravitational Asymptotic Safety program \cite{Percacci:2017fkn,Reuter:2019byg} strives for a quantum theory of gravity valid on all scales. The construction is not limited to pure gravity and can be supplemented by a broad class of matter degrees of freedom \cite{Eichhorn:2018yfc}. As its characteristic feature, the program builds on well-established principles of relativistic quantum field theory (QFT), and postulates that the dynamics of gravity in the ultraviolet (UV) is controlled by an interacting renormalization group (RG) fixed point \cite{Weinberg:1980gg,Reuter:1996cp}. In contrast to theories enjoying asymptotic freedom, interactions are not turned off in this regime though. Instead, classical and quantum contributions balance in a delicate way such that the high-energy regime of the theory enjoys an additional symmetry, so-called quantum scale invariance \cite{Wetterich:2019qzx}. It is expected that this feature entails the absence of unphysical UV divergences in physical observables. The ambition of the \emph{form factor framework to asymptotically safe quantum gravity}, initiated in \cite{Knorr:2019atm}, is to make this expectation precise both at the level of scattering amplitudes, and at the level of quantum corrections to spacetimes exhibiting singularities at the classical level.
	
	The pivotal element in the form factor framework is the quantum effective action $\Gamma$. By definition, the propagators and vertices derived from this generating functional are exact in the sense that they include all quantum corrections. Thus, quantum-corrected spacetimes may be constructed by solving the quantum equations of motion derived from $\Gamma$. Moreover, quantum-corrected scattering processes are described by tree-level Feynman diagrams built from the effective propagators and vertices provided by the quantum effective action. Thus the computation of $\Gamma$ is considered as equivalent to solving the theory. Clearly, obtaining the quantum effective action from a first principle computation is then a notoriously hard problem. This applies in particular in the context of gravity, where the long-range nature of the gravitational force introduces non-local interaction terms at the effective level. 
	
	Given this rather intimidating perspective, it is useful to break the analysis of $\Gamma$ into several parts. Firstly, one would like to have a conceptual understanding of the building blocks that are essential for implementing the concept of Asymptotic Safety at the level of the quantum effective action. This leads to the form factors introduced in \autoref{sec.ea}. The characteristic feature of the form factors is that they encode the full momentum dependence of propagators and vertices in the context of a general spacetime. This makes the form factors an indispensable element for encoding the gravitational dynamics taking quantum corrections into account. 
	
	The next step seeks to understand the implications of form factors at the level of physical processes. A prime example is the scattering of particles, where the amplitudes encoding the probability of a process receive non-trivial contributions from the form factors. The general structure of these amplitudes can be understood systematically by identifying all terms in $\Gamma$ that contribute to a given scattering process. For a particular process with finitely many external particles, this classification involves only a finite number of form factors. In particular, it is not necessary to determine the quantum effective action in full generality. Once this classification is completed, one derives the most general amplitude compatible with the existence of a quantum effective action. The results obtained along these lines are summarized in \autoref{sec.scattering} where we explain the role of form factors in gravity-mediated two-to-two scattering processes with external matter fields.
	
	Based on this parametric approach, it is then natural to ask about the conditions on the form factors leading to \emph{asymptotically safe scattering amplitudes}. A proof of principle demonstrating that the form factors indeed provide sufficient room to implement this scenario is provided in \autoref{sec.tanh}. The key insights from this specific example are that asymptotically safe amplitudes may be obtained without introducing new degrees of freedom (akin to string theory): the specific example tames the growth of the amplitude at high energy through a Regge-like structure of first-order poles situated at purely imaginary squared momenta. The model also demonstrates that Asymptotic Safety at the level of amplitudes requires a delicate interplay between the momentum dependence in the propagators and vertices. While this ``tuning'' may appear to be ad hoc at first sight, it is natural to expect that these structures will actually be provided by the interacting RG fixed point underlying Asymptotic Safety.
	
	A key question within the Asymptotic Safety program is whether the dynamics resulting from the RG fixed point is compatible with the principles of causality, unitarity, and positivity bounds formulated at the level of the $S$-matrix. Important cross-checks in this direction arise from extracting the low-energy effective field theory from $\Gamma$ by expanding the interactions in inverse powers of a UV cutoff scale, and truncating the expansion at a fixed order. Based on low-energy effective field theory considerations, one can then derive consistency conditions on the couplings appearing in this expansion. Prominent examples are the constraints on the graviton three-point vertex derived by Maldacena et.\ al.\ \cite{Camanho:2014apa} as well as restrictions on the signs of the matter self-interactions, see \eg{}, \ \cite{Cheung:2014ega,Tokuda:2020mlf,Alberte:2020bdz,Alberte:2021dnj}. \autoref{sec.eft} gives a brief overview on such consistency checks. The typical assumptions made in their derivation are of relevance in the context of Asymptotic Safety.
	
	The relevance of the form factor framework for implementing Asymptotic Safety clearly warrants the derivation of these functions from first principle computations. \autoref{sec.hardcore} briefly covers the two strategies that have been employed in this context: solutions of the functional RG, foremost the Wetterich equation \cite{Wetterich:1992yh,Morris:1993qb,Reuter:1996ub}, and the reconstruction of form factors based on correlation functions obtained from Monte Carlo simulations. While this program is still in its infancy, it provided some interesting pointers to the presence of non-local form factors which could lead to phenomenologically interesting modifications of the gravitational dynamics on macroscopic scales \cite{Knorr:2018kog}.
	
	The main focus of this chapter is on the role of form factors within the gravitational Asymptotic Safety program. As stressed in our conclusions, \autoref{sec.conclusion}, the application range of this framework is actually much broader. In particular, the form factors readily capture quantum corrections arising within the effective field theory of quantum gravity \cite{Donoghue:2017pgk}, non-local ghost-free gravity \cite{Biswas:2005qr,Biswas:2011ar,Buoninfante:2018xiw,Buoninfante:2020ctr}, and perturbatively renormalizable approaches to quantum gravity \cite{Modesto:2014lga,Modesto:2017hzl,Modesto:2017sdr}. Therefore, the form factor framework has the potential of providing a unifying perspective on quantum gravity and its phenomenological consequences.

	\section{The quantum effective action including form factors}
		\label{sec.ea}
Form factors are the key element for constructing (gravity-mediated) scattering amplitudes that are well-behaved at trans-Planckian energy scales. This section introduces the concept of form factors for the effective action $\Gamma$ built from a Lorentzian spacetime metric $g_{\mu\nu}$, an Abelian gauge field $A_\mu$, an uncharged scalar field $\phi$, and Dirac fermions $\psi$. In particular, we present the results from the classification program \cite{Knorr:2019atm,Draper:2020bop,Draper:2020knh,Knorr:2021iwv,Knorr:2022lzn}, constructing all interaction monomials which contribute to the two-to-two scattering of matter fields in a flat spacetime.

	\subsection{Setup}
	
	Here we collect the conventions used in the description of the effective action. In particular, we present our notation for spacetime fields and curvature, and underlying symmetry assumptions. We also introduce the notation used to efficiently denote form factors.
	\subsubsection{Conventions: fields and Bianchi identities}
	We work on a generic four-dimensional spacetime with metric $g_{\mu\nu}$ and signature $\{+,-,-,-\}$. The covariant derivative associated with this metric is  denoted by $\CD_\mu$, and we introduce the covariant d'Alembertian $\dalembertian \equiv - g^{\mu\nu} \CD_\mu \CD_\nu$. In order to ease our notation, the contraction of spacetime indices will frequently be denoted by ``$\cdot$'', \eg{}, $\dalembertian = - \CD \cdot \CD$. We define the Riemann tensor as
	\begin{equation}\label{def.riemann}
	R_{\mu\nu\rho}{}^\lambda \equiv \p_\nu \Gamma^\lambda{}_{\mu\rho} - \p_\mu \Gamma^\lambda{}_{\nu\rho} + \Gamma^\sigma{}_{\mu\rho} \Gamma^\lambda{}_{\nu\sigma} + \Gamma^\sigma{}_{\nu\rho} \Gamma^\lambda{}_{\mu\sigma} \, ,
	\end{equation}
	and the Ricci tensor and Ricci scalar are $R_{\mu\nu} = R_{\mu\lambda\nu}{}^\lambda$ and $R = g^{\mu\nu} R_{\mu\nu}$. The Riemann tensor satisfies the Bianchi identities
	\begin{equation}\label{eq.bianchiR}
	R_{\mu[\nu\rho\sigma]} = 0 \, , \qquad \CD_{[\alpha} R_{\mu\nu]\rho\sigma} = 0 \, . 
	\end{equation}
	Here $[\cdots]$ denotes anti-symmetrization with unit strength. The second Bianchi identity implies the contracted Bianchi identities
	\begin{equation}\label{eq.biancicontracted}
	\CD^\alpha R_{\alpha\beta \mu\nu} = 2 \CD_{[\mu} R_{\nu]\beta} \, , \qquad \CD^\nu R_{\mu\nu} = \frac{1}{2}\CD_\mu R \, . 
	\end{equation}
	As a direct consequence, one has that 
	\begin{equation}\label{eq.mapffbasic}
	\dalembertian R_{\rho\sigma\mu\nu} = 2 \CD_\sigma \CD_{[\mu} R_{\nu]\rho} - 2 \CD_\rho \CD_{[\mu} R_{\nu]\sigma} + O(R^2) \, . 
	\end{equation}
	Contracting the open indices with the Riemann tensor and integrating by parts then yields the relation \cite{Codello:2012kq,Knorr:2019atm}
	\begin{equation}\label{mapff}
	\int \dd^4x \sqrt{-g} \left\{R^{\rho\sigma\mu\nu} \, \dalembertian^n \,  R_{\rho\sigma\mu\nu} - 4 R^{\mu\nu} \, \dalembertian^n \,  R_{\mu\nu} + R \,\dalembertian^n \,  R \right\} = O(R^3) \, , \quad n \ge 1 \, , 
	\end{equation}
	where the right-hand side denotes terms which are of third order in the spacetime curvature tensors. In four dimensions, this identity is complemented by the Gauss-Bonnet identity, stating that the combination 
	\begin{equation}\label{eq.gaussbonnet}
	E = R^2 - 4 R_{\mu\nu} R^{\mu\nu} + R_{\mu\nu\rho\sigma} R^{\mu\nu\rho\sigma}
	\end{equation}
	is topological in the sense that it does not contribute to the equations of motion. Finally, it is convenient to express the Riemann tensor in terms of the Weyl tensor
	\begin{equation}\label{eq.weyl}
	C_{\mu\nu\rho\sigma} = R_{\mu\nu\rho\sigma} - g_{\mu[\rho} R_{\sigma]\nu} +  g_{\nu[\rho} R_{\sigma]\mu} + \frac{1}{3} R g_{\mu[\rho} g_{\sigma]\nu} \, . 
	\end{equation}

In the gauge sector, we introduce the gauge-invariant Abelian field strength tensor of the photon $A_\mu$,
\begin{equation}\label{eq.fieldstrength}
	F_{\mu\nu} = \partial_\mu A_\nu - \partial_\nu A_\mu
	\,\text{.}
\end{equation}
 The field strength tensor satisfies a Bianchi identity
\begin{equation}\label{eq.bianchiF}
	 \CD_{[\alpha} F_{\beta\gamma]} = 0
	\,\text{.}
\end{equation}

In the fermionic sector, we introduce covariant Dirac matrices $\{\gamma_\mu\}$, that satisfy the anti-commutator
\begin{equation}\label{eq.Clifford}
	\{\gamma_\mu,\gamma_\nu\} = 2 g_{\mu\nu} \mathbb{1}
	\,\text{.}
\end{equation}
This gives the Dirac operator $\slashed{\CD} = g^{\mu\nu} \gamma_\mu \nabla_\nu$, which satisfies the Lichnerowicz relation:
\begin{equation}\label{eq.Lichnerowicz}
	\Delta_{\text{D}}	\equiv	(\imath \slashed \CD)^2	=	\left(	\dalembertian	+	\frac{1}{4}	R	\right)	\mathbb{1}
	\,\text{.}
\end{equation}
Furthermore, we have the matrix
\begin{equation}\label{eq.gamma5}
	\gamma_\star = \frac{1}{24}	\sqrt{-g}	\epsilon^{\mu\nu\rho\sigma}	\gamma_\mu\gamma_\nu\gamma_\rho\gamma_\sigma
	\,\text{.}
\end{equation}
We use the symbols $\Phi$ and $\Psi$ to denote generic fields which can either be metric fluctuations, gauge, or matter fields.

\subsubsection{Form factors}
The key idea of a form factor is to promote the coupling constants associated with a given interaction term to a momentum-dependent function. A prototypical example is provided by the electric charge $e$ appearing in electrodynamics,
\begin{equation}\label{eq.electricexample}
\frac{1}{4} \int \dd^4x \sqrt{-g} F_{\mu\nu} \frac{1}{e^2} F^{\mu\nu} \quad \mapsto \quad
\frac{1}{4} \int \dd^4x \sqrt{-g} F_{\mu\nu} \frac{1}{e^2(\dalembertian)} F^{\mu\nu} \, , 
\end{equation}
where $e^2$ is promoted to a function $e^2(\dalembertian)$. \emph{In this way the (corrections to) the dynamics is encoded in a manifestly background-independent way.} In Minkowski spacetime, the $e^2(\dalembertian)$ reduces to a momentum-dependent function via Fourier transformation. In the case of curved spacetime, this is straightforwardly generalized by replacing partial derivatives by covariant ones. In this way, the form factors capture the well-known momentum dependence (colloquially also called the running) of couplings observed in particle physics experiments \cite{ParticleDataGroup:2016lqr} at the level of the effective action. The form factor framework implements this generalization in a systematic way.

Generically, the form factors depend on all independent contractions of the covariant derivatives. By partial integration, we can reduce the number of arguments of the form factors. We adopt the following notation: for a form factor acting on $\Phi_1 \cdots \Phi_n$, the arguments are written as
\begin{align}\label{eq.arguments23}
	\int \dd^4x \sqrt{-g} \, \Phi_1 f(\dalembertian)	\Phi_2
	\,\text{,}&\qquad&
	\int \dd^4x \sqrt{-g} \, f(\dalembertian_1,\dalembertian_2,\dalembertian_3)	\Phi_1 \Phi_2 \Phi_3
	\,\text{,}
\end{align}
and
\begin{align}\label{eq.argumentsn}
	\int \dd^4x \sqrt{-g} \, f\left(	\{-\CD_i \cdot\CD_j\}_{1\leq i < j \leq n}	\right)	\Phi_1 \cdots	\Phi_n
	\,\text{,}&\qquad& 
	n\geq 4
	\,\text{.}
\end{align}
In particular, form factors associated with monomials containing four fields generically have six independent arguments. The subscript on each operator denotes the field it acts on, for example $\CD_1(\Phi_1 \Phi_2) = (\CD_1 \Phi_1) \Phi_2$. In order to ease the notation, we will often suppress the arguments of the form factors.

\subsubsection{Symmetry assumptions}
We constrain the terms tracked in our classification by imposing the following symmetry requirements.  The gauge field $A_\mu$ comes with a $\mathrm{U}(1)$ gauge symmetry. Therefore, the dependence of $\Gamma$ on the photon must be in the form of the field strength tensor \eqref{eq.fieldstrength}. We will assume that both scalars and fermions are uncharged under $\mathrm{U}(1)$. Furthermore, we impose that each scalar field comes with a global $\mathbb{Z}_2$-symmetry, so that only even powers appear in the effective action. In addition, the complete action is invariant under diffeomorphism symmetry.
	\subsection{Classifying the interactions within the effective action}
We are interested in finding all interaction monomials that can contribute to the graviton-mediated two-to-two particle scattering process. Since the $n$-point function is obtained by taking $n$ functional derivatives, it is in general determined by terms in $\Gamma$ containing $n$ fields. This motivates the notation of $\Gamma_{\Phi^n \Psi^m}$ to denote the building block of $\Gamma$ containing $n$ fields of type $\Phi$ and $m$ fields of type $\Psi$.

Gravitons (and non-Abelian gauge fields) form a notable exception to this classification. Since each term in the effective action will contain a factor $\sqrt{-g}$, any term in $\Gamma$ will be non-linearly coupled to gravity. Moreover, spacetime curvature tensors contain infinitely many powers of the metric fluctuations. However, since we are expanding around a flat background, any term containing more than $n$ curvature tensors will not contribute to a vertex with $n$ graviton legs. Therefore, we adopt the convention that  $\Gamma_{h^n}$ contains up to $n$ curvature tensors, while for $m\geq 1$, the building block $\Gamma_{h^n \Phi^m}$ contains exactly $n$ curvature tensors and $m$ fields of type $\Phi$.
\subsubsection{Pure gravity}
Let us start by discussing the gravitational sector. Thinking about gravitons $h_{\mu\nu}$ propagating in a flat Minkowski metric $\eta_{\mu\nu}$, it is natural to organize the effective action in terms of an expansion in powers of the spacetime curvature. Neglecting terms which are cubic in the Riemann tensor or its contractions, the most general form of the effective action is given by
	\begin{equation}\label{eq.Gammah2}
		\Gamma_{h^2}	=	\frac{1}{16\pi \GN}	\int \dd^4x\sqrt{-g}	\, \left[2 \Lambda	-R	-	\frac{1}{6}	R f_{RR}(\dalembertian)R	+	\frac{1}{2}	C_{\mu\nu\rho\sigma}	f_{CC}(\dalembertian)	C^{\mu\nu\rho\sigma}	\right]
		\,\text{.}
	\end{equation}
Here $\GN$ is Newton's constant, $\Lambda$ denotes the cosmological constant, and $f_{RR}(\dalembertian)$ and $f_{CC}(\dalembertian)$ are the two form factors appearing at quadratic order in the spacetime curvature. We stress that these couplings are effective in the sense that they incorporate all quantum corrections. This also entails that $\GN$ and $\Lambda$ are constant \cite{Knorr:2019atm,Bonanno:2020bil}. Similarly, \eqref{eq.Gammah2} entails that the couplings appearing in the quadratic part of the action can develop a momentum dependence which is encoded in the corresponding form factors.

The expansion \eqref{eq.Gammah2} is complete in the sense that there is no contribution associated with the square of the Ricci tensor. Any contributions containing the d'Alembertian can be eliminated via the identity \eqref{mapff}, while the pure $R_{\mu\nu} R^{\mu\nu}$-term can be rewritten in terms of the Gauss-Bonnet combination \eqref{eq.gaussbonnet}, and then does not enter physical processes. When writing \eqref{eq.Gammah2}, we adopted the ``Weyl-basis'' which leads to an easy relation between the form factors $f_{RR}$, $f_{CC}$ and the graviton propagator. The identities \eqref{eq.weyl} and \eqref{mapff} allow to map this result to the Ricci-basis where the form factors are associated with the squares of the Ricci scalar and Ricci tensor.

While \eqref{eq.Gammah2} suffices to determine the graviton propagator in flat space, studying two-to-two graviton scattering requires extending this result to quartic order in the spacetime curvature. Building on the FKWC-classification \cite{Fulling:1992vm}, a series of geometric identities valid at cubic order has been published in \cite{Decanini:2008pr}. A local basis for invariants at cubic order in the spacetime curvature can be found in \cite{Reuter:2019byg} and an extension to fourth-order, tailored to graviton scattering, has been pursued in \cite{Chowdhury:2019kaq}, see also \cite{Knorr:2020ckv} for a basis given in the context of the functional RG. Owed to the Bianchi identity, the generalization of these results including form factors is a formidable task. Some systematics in a \emph{non-local} basis have been developed in the context of heat kernel computations \cite{Barvinsky:1990up, Barvinsky:1993en}.

	\subsubsection{Gravity coupled to scalar matter}
	We proceed by considering gravity coupled to an uncharged, massive real scalar field $\phi$. Following the example \eqref{eq.electricexample}, the kinetic term including the form factor is
	\begin{equation}\label{eq.Gammaphi2}
		\Gamma_{\phi^2}	=	\frac{1}{2} \int \dd^4 x\sqrt{-g} \,	\phi	f_{\phi\phi}(\dalembertian)	\phi
		\,\text{.}
	\end{equation}
	The form factor is normalized such that $f^\prime_{\phi\phi}(m^2) = 1$ which ensures that the scalar field is canonically normalized on-shell.
	
	The construction of a basis for the interaction vertices has to account for the redundancies due to partial integration. In flat space, this amounts to rewriting the momentum dependence of the vertices by exploiting momentum conservation. A basis for the graviton-scalar-scalar vertex is then provided by
	\begin{equation}\label{eq.Gammahphi2}
			\Gamma_{h\phi^2}	
		=
			\int \dd^4 x \sqrt{-g}\,	\bigg[
				f_{R\phi\phi}	R\phi\phi	
			+	f_{\text{Ric}\phi\phi}	R^{\mu\nu}	(\CD_\mu\phi)(\CD_\nu\phi)	
		\bigg]
		\,\text{.}
	\end{equation}
	The four-point vertex for the scalar self-interaction is created by a single term,
	\begin{equation}\label{eq.Gammaphi4}
			\Gamma_{\phi^4}
		=
			\int \dd^4 x\sqrt{-g}	\,	f_{\phi^4}	\,	\phi	\phi	\phi	\phi
		\,\text{.}
	\end{equation}
In the latter case, all derivatives acting on the scalars are created through the form factor. This is readily seen by noting that the six independent arguments in $f_{\phi^4}$ suffice to generate all possible contracted derivatives acting on the three left-most $\phi$-fields. At the same time any derivative acting on the fourth scalar field can always be removed by partial integration.

At this stage it is instructive to provide an explicit example on how non-basis monomials are mapped to basis elements. For explicitness, we consider a contribution to the $h\phi\phi$-vertex of the form
\begin{equation}\label{eq.scalarexample}
I = \int \dd^4x \sqrt{-g} \, f_{R\phi\phi}(\dalembertian_1,\dalembertian_2,\dalembertian_3) \,	R \, (\CD_\mu\phi)(\CD^\mu\phi) \, . 	
\end{equation}
The symmetry in the scalar fields ensures that $f_{R\phi\phi}$ is symmetric in its last two arguments. In order to map $I$ to the basis \eqref{eq.Gammahphi2}, we rewrite the term as
\begin{equation}
\begin{aligned}
 I &= -\frac{1}{2} \int \dd^4x \sqrt{-g} \, f_{R\phi\phi}(\dalembertian_1,\dalembertian_2,\dalembertian_3) \, R \, \left[ \dalembertian \left( \phi\phi \right) - \left( \dalembertian \phi \right) \phi - \phi \left( \dalembertian \phi \right) \right] \\
 &= -  \frac{1}{2} \int \dd^4x \sqrt{-g} \,  f_{R\phi\phi}(\dalembertian_1,\dalembertian_2,\dalembertian_3) \, \left[  \, \dalembertian_1 -  \dalembertian_2 -  \dalembertian_3 \right] \, R \phi\phi \, .
\end{aligned}
\end{equation}
The last line matches the structure of the first term in \eqref{eq.Gammahphi2}. Thus at this point \eqref{eq.scalarexample} has been mapped to the basis provided in \eqref{eq.Gammahphi2}.
	\subsubsection{Gravity coupled to photons}
The classification of interactions among gravitons and photons follows along similar lines as the scalar case. Introducing a form factor, the photon kinetic term becomes
	\begin{equation}\label{eq.GammaF2}
		\Gamma_{A^2}	=	- \frac{1}{4}	\int \dd^4x\sqrt{-g}\,	F_{\mu\nu} f_{FF}(\dalembertian) F^{\mu\nu}
		\,\text{.}
	\end{equation}
The overall normalization of $f_{FF}(\dalembertian)$ is again fixed by imposing that the field is canonically normalized on-shell: $f_{FF}(0) = 1$.

The construction of a basis for the interaction vertices including the appropriate form factors is slightly more involved than in the scalar case, since the photon case has three operations which allow to map interaction monomials to each other. Firstly, there is partial integration. Secondly, one has to account for the anti-symmetry of the field strength $F_{\mu\nu}$. Thirdly, derivatives acting on the field strength can be rewritten by applying the Bianchi identity \eqref{eq.bianchiF}. Following \cite{Knorr:2022lzn}, a basis can be identified based on the following algorithm: one starts with the product of a fixed number of uncontracted field strength tensors and covariant derivatives.
Subsequently, one constructs the highly over-complete set of interaction monomials by performing all possible contractions of the spacetime indices. Finally, this set is reduced to the minimal number of independent terms by applying all symmetry operations specified above in a systematic way.

For the graviton-photon-photon vertex extracted from interaction monomials with one spacetime curvature, this procedure identifies seven independent form factors. These can be chosen according to
	\begin{equation}\label{eq.GammahA2}
		\begin{aligned}
		\Gamma_{hA^2}	&=	\int	\dd^4 x	\sqrt{-g}	\,	\bigg[	
			f_{RFF}				\,	R	F_{\alpha\beta}	F^{\alpha\beta}
			+	f_{\text{Ric}FF}	\,	R^{\alpha\beta}	\tensor{F}{_\alpha^\gamma}	\tensor{F}{_\beta_\gamma}
			\\&\qquad
			+	f_{\text{Rm}FF}	\,	R_{\alpha\beta\gamma\delta}	F^{\alpha\beta}	F^{\gamma\delta}
			\\&\qquad
			+	f_{D^2RFF}				\,	(\CD^\alpha \CD^\beta R)	\tensor{F}{_\alpha^\gamma}	\tensor{F}{_\beta_\gamma}
			+	f_{D^2\text{Ric}FF}	\,	(\CD^\alpha	\CD^\beta	R^{\gamma\delta})	F_{\alpha\gamma}	F_{\beta\delta}
			\\&\qquad
			+	f_{\text{Ric}D^2FF}	\,	R^{\gamma\delta}	(\CD^\alpha	\CD^\beta	F_{\alpha\gamma})	F_{\beta\delta}
			+	f_{\text{Ric}DFDF}	\,	R^{\gamma\delta}	(\CD^\alpha	F_{\alpha\gamma})	(\CD^\beta	F_{\beta\delta})
			\bigg]
			\,\text{.}
		\end{aligned}
	\end{equation}
The identity \eqref{eq.mapffbasic} then leads to constraints on the functional form of the form factors. For instance $f_{\text{Rm}FF}$ is independent of $\dalembertian_1$ since the Bianchi identity satisfied by the Riemann tensor allows to map terms of the form $\dalembertian R_{\rho\sigma\mu\nu}$ to other basis elements in \eqref{eq.GammahA2}. In a non-local basis, this term can be removed entirely.

Similarly, one concludes that the four-photon self-interactions contain seven free functions that can be chosen according to
\begin{equation}\label{eq.GammaA4}
	\begin{aligned}\Gamma_{A^4}
		&=
		\int	\dd^4 x	\sqrt{-g}\, \bigg[
		f_{F^2F^2}	\,	F_{\alpha\beta}	F^{\alpha\beta}	F_{\gamma\delta}	F^{\gamma\delta}
		+	f_{F^4}		\,	\tensor{F}{_\alpha^\beta}	\tensor{F}{_\beta^\gamma}	\tensor{F}{_\gamma^\delta}	\tensor{F}{_\delta^\alpha}	
		\\&\qquad
		+	f_{FFDFDF_1}	\,	\tensor{F}{_\alpha^\gamma}	F^{\delta\zeta}	(\CD^\alpha	F_{\beta\delta})	(\CD^\beta	F_{\gamma\zeta})
		+	f_{FFDFDF_2}	\,	\tensor{F}{_\beta^\gamma}	F^{\delta\zeta}	(\CD^\alpha	F_{\alpha\gamma})	(\CD^\beta	F_{\delta\zeta})
		\\&\qquad
		+	f_{FFDFDF_3}	\,	F_{\alpha\beta}	F^{\gamma\delta}	(\CD^\alpha	\tensor{F}{_\gamma^\zeta})	(\CD^\beta	F_{\delta\zeta})
		+	f_{FFDFDF_4}	\,	\tensor{F}{_\beta^\gamma}	F_{\alpha\gamma}	(\CD^\alpha	F^{\delta\zeta})	(\CD^\beta	F_{\delta\zeta})
		\\&\qquad
		+	f_{FFD^2FD^2F}	\,	F_{\alpha\gamma}	F_{\beta\delta}	(\CD^\alpha	\CD^\beta	F^{\zeta\kappa})	(\CD^\gamma \CD^\delta	F_{\zeta\kappa})
		\bigg]
		\,\text{.}\end{aligned}	
\end{equation}
We note that the first line can also be formulated in terms of the dual field strength $\tilde{F}_{\mu\nu} \equiv \frac{1}{2} \epsilon_{\mu\nu\rho\sigma} F^{\rho\sigma}$. Applying standard identities for products of totally anti-symmetric tensors, one has
\begin{equation}\label{eq.dualfield}
2 \left( F_{\mu\nu} F^{\mu\nu} \right)^2 + \left( F_{\mu\nu} \tilde{F}^{\mu\nu} \right)^2 =  4 \,  \tensor{F}{_\alpha^\beta}	\tensor{F}{_\beta^\gamma}	\tensor{F}{_\gamma^\delta}	\tensor{F}{_\delta^\alpha}	\, . 
\end{equation}
This relates the basis used in \eqref{eq.GammaA4} to the standard formulation of the Euler-Heisenberg Lagrangian.

Remarkably, the inclusion of form factors also leads to new interactions that otherwise would vanish due to the anti-symmetry of the field strength. In particular, there are two form factors associated with the three-photon vertex
	\begin{equation}\label{eq.GammaA3}
			\Gamma_{A^3} =	\int \dd^4x \sqrt{-g}	\,	\left[
				f_{F^3}	\,	\tensor{F}{_\alpha^\beta}	\tensor{F}{_\beta^\gamma}	\tensor{F}{_\gamma^\alpha}
			+	f_{FDFDF}	\,	\tensor{F}{^\gamma^\delta}	(\CD^\alpha\tensor{F}{_\alpha_\gamma})	(\CD^\beta\tensor{F}{_\beta_\delta})
			\right]
			\,\text{.}
	\end{equation}
We stress that while one can clearly write down these terms as part of the classification program, this does not necessarily entail that the corresponding form factors will actually appear in the effective action. Their appearance can be obstructed by global symmetries as, \eg{}, invariance of the action under $F_{\mu\nu} \mapsto - F_{\mu\nu}$ which is respected by \eqref{eq.GammahA2} and \eqref{eq.GammaA4} but broken in \eqref{eq.GammaA3}.

\subsubsection{Matter self-interactions coupling scalars and photons}
As a side-product of the classification described in the previous subsection, one can also determine a basis for scalar-photon-interactions, including form factors. For two scalars and one photon, we have
\begin{equation}\label{eq.GammaAphi2}
	\Gamma_{A\phi^2}	=	\int \dd^4 x \sqrt{-g} \, f_{F\phi^2}	F^{\alpha\beta}	(\CD_\alpha \phi)	(\CD_\beta \phi)
	\,\text{,}
\end{equation}
where $f_{F\phi^2}(\dalembertian_1,\dalembertian_2,\dalembertian_3)$ must be anti-symmetric in the last two arguments to account for the anti-symmetry of the field strength tensor. Finally, the two-scalar-two-photon vertex is fixed by
	\begin{equation} \label{eq.GammaA2phi2}
		\begin{aligned}	\Gamma_{A^2\phi^2}
			&=
			\int	\dd^4 x	\sqrt{-g}	\,	\bigg[
			f_{FF\phi^2}	\,	F_{\alpha\beta}	F^{\alpha\beta}	\phi \phi
			+	f_{FFD\phi D\phi}		\,	\tensor{F}{_\alpha^\gamma}	\tensor{F}{_\gamma_\beta}	(\CD^\alpha \phi)	(\CD^\beta \phi)
		\\&\quad
			+	f_{FF D^2\phi \phi}	\,	\tensor{F}{_\alpha^\gamma}	\tensor{F}{_\gamma_\beta}	(\CD^\alpha \CD^\beta	\phi)	\phi
			+	f_{FFD^2\phi D^2\phi}	F_{\alpha\gamma}	F_{\beta\delta}	(\CD^\alpha \CD^\beta\phi)	(\CD^\gamma \CD^\delta	\phi)
			\bigg]
			\,\text{.}\end{aligned}
\end{equation}
	\subsubsection{Fermions}
The form factor framework is readily extended to Dirac spinors $\psi$, but it has not been carried out yet to a similar degree as for the other non-gravitational fields. In this context, we first note that the Dirac operator satisfies the Lichnerowicz relation \eqref{eq.Lichnerowicz}. Thus, it is natural to choose the differential operator appearing as the argument of the form factor as $\Delta_{\text{D}} \equiv (\imath \slashed{\CD})^2$. This entails that the form factors actually commute with the Dirac operator. In a similar fashion, one would use the gauge-covariant derivative for fields charged under a gauge symmetry.

For uncharged fermions and up to quadratic order in $\psi$, we then encounter four form factors, one associated with the kinetic term and mass term of each chirality. Thus we can form the following bi-linears \cite{Knorr:2019atm}:
	\begin{equation}\label{eq.Gammapsi2}
	\begin{aligned}
		\Gamma_{\bar\psi\psi}	&=	\int \dd^4x \sqrt{-g}	\bar \psi\Big(
			f_{\bar\psi\psi,1}(\Delta_{\text{D}})	(\imath \slashed{\CD})
		+	f_{\bar\psi\psi,2}(\Delta_{\text{D}})	\gamma_\star	(\imath \slashed{\CD})
		\\&\hspace{4cm}
		+	f_{\bar\psi\psi,3}(\Delta_{\text{D}})	\mathbb{1}
		+	f_{\bar\psi\psi,4}(\Delta_{\text{D}})	\gamma_\star
		\Big)	\psi
		\,\text{.}
	\end{aligned}
	\end{equation}
This concludes our classification of interaction monomials which can appear in the effective action. As we will see in \autoref{sec.scattering}, the results are sufficient to describe the most general gravity-mediated two-to-two scattering amplitudes including scalars and photons as external particles.
\subsection{Field re-definitions and inessential operators}\label{sec:fieldredef} 
We conclude our discussion of parameterizations of the effective action with two important technical remarks. Starting from the scalar kinetic term \eqref{eq.Gammaphi2}, it is tempting to write
\begin{equation}\label{eq.fdecomposition}
f_{\phi\phi}(\dalembertian) = (\dalembertian - m^2) \tilde{f}_{\phi\phi}(\dalembertian) 
\end{equation}
and to subsequently absorb the factor $\tilde{f}_{\phi\phi}(\dalembertian)$ in a momentum-dependent field redefinition
\begin{equation}\label{eq.fieldredef}
\phi \mapsto \tilde{\phi} \equiv (\tilde{f}_{\phi\phi}(\dalembertian))^{1/2} \phi \, . 
\end{equation}
This would remove the form factor from the kinetic term at the expense of modifying the form factors encoding the moment-dependent interaction. Clearly, physics should not be affected by this redefinition. The decomposition \eqref{eq.fdecomposition} followed by \eqref{eq.fieldredef} presupposes that we have identified the degrees of freedom of the theory (encoded in the poles of the scalar propagator) to be the ones of a massive scalar field with mass $m$. In this case, $\tilde{f}_{\phi\phi}(\dalembertian)$ will be a positive and invertible function and the field redefinition is well-defined.  This logic fails, however, if the theory contains additional degrees of freedom inducing zeros in $\tilde{f}_{\phi\phi}(\dalembertian)$. In this case, the field redefinition \eqref{eq.fieldredef} would be ill-defined. Heuristically, this is easily understood from the observation that the theories before and after the field redefinition would differ in their degrees of freedom. In order to be as general as possible, we therefore retain the form factors in the kinetic terms in our classification. 

More generally, our classification does not include information about interactions which can be removed by field redefinitions -- called redundant (or inessential) operators \cite{Wegner_1974, Hawking:1979ig, Dietz:2013sba, Baldazzi:2021ydj, Baldazzi:2021orb}. By definition, a redundant operator does not contribute to observables. It is a scaling operator with respect to the RG which vanishes upon imposing the equations of motion. The second property implies that it can be written as
\begin{equation}\label{red.op}
\cO = \int \dd^4x \sqrt{-g} \left[ \frac{1}{\sqrt{-g}} \frac{\delta \Gamma}{\delta \Phi} \right] \, \cdot \, F[\Phi] \, . 
\end{equation}
The functional $F[\Phi]$ is related to an infinitesimal field redefinition
\begin{equation}
\Phi \mapsto \Phi + \epsilon \, F[\Phi] \, ,
\end{equation}
and can depend on the spacetime coordinates $x^\mu$ as well as on the fields $\Phi$ and their derivatives.

The key property of \eqref{red.op} is that the definition of a redundant operator requires knowledge about the equations of motion. It is instructive to illustrate this property in the context of gravity. For simplicity, we take the scaling properties to be the one of the Gaussian fixed point, \ie, all operators scale according to their classical mass dimension. Suppose $\Gamma$ is given by the Einstein-Hilbert action without cosmological constant,
\begin{equation}\label{eq.eh}
\Gamma[g] = - \frac{1}{16\pi \GN} \int \dd^4x \sqrt{-g} \, R \, . 
\end{equation}
The resulting equations of motion are
\begin{equation}\label{eq.einstein}
R_{\mu\nu} - \frac{1}{2} g_{\mu\nu} R = 0 \, . 
\end{equation}
Taking $F_{\mu\nu}[g] = a g_{\mu\nu} R + b R_{\mu\nu}$ in \eqref{red.op} then yields
\begin{equation}\label{red.eh}
\cO = \int \dd^4x \sqrt{-g} \left[ b R_{\mu\nu} R^{\mu\nu} - \frac{1}{2} (2a+b) R^2 \right] \, . 
\end{equation}
Hence we find that two of the three possible quadratic curvature terms are redundant. In combination with the property that the third curvature-squared term can be written in terms of the topological Gauss-Bonnet integrand, this underlies the well-known result that the perturbative quantization of the Einstein-Hilbert action \eqref{eq.eh} does not require a counterterm at the one-loop level \cite{'tHooft:1974bx}. At sixth order in derivatives, the same procedure allows to eliminate all invariants except the one cubic in the Weyl tensor. It is exactly the invariant that appears at two loops in perturbation theory \cite{Goroff:1985sz, Goroff:1985th}.

It is instructive to repeat this analysis for the case where $\Gamma$ is given by the action for quadratic gravity. Written in the Ricci-Weyl basis, we then have
\begin{equation}\label{eq.qg}
\Gamma[g] = \int \dd^4x \sqrt{-g} \left[ \frac{1}{2} \alpha \, C_{\mu\nu\rho\sigma} C^{\mu\nu\rho\sigma}  - \frac{1}{6} \beta R^2 \right] \, . 
\end{equation}
The resulting equations of motion are
\begin{equation}\label{eom.qg}
		\alpha \left( \CD^\rho \CD^\sigma + \frac{1}{2} R^{\rho\sigma} \right) C_{\mu\rho\nu\sigma} 
	-	\frac{\beta}{6} \left( R_{\mu\nu} - \frac{1}{4} g_{\mu\nu}R - \CD_\mu \CD_\nu + g_{\mu\nu} \CD^2 \right) R = 0 \, . 
\end{equation}
To remove inessential operators at sixth order in derivatives, we again use $F_{\mu\nu}[g] = a g_{\mu\nu} R + b R_{\mu\nu}$. Evaluating \eqref{red.op} gives a linear combination of several monomials, and by choosing $a$ and $b$, two of them can be made redundant. In quadratic gravity, we are thus left with six relevant operators at this order, compared to only one starting from an Einstein-Hilbert theory.

In this light, the classification presented in this section does not reference a specific dynamics. In other words, there was no attempt to identify the redundant operators in our basis, since the definition of redundant operators hinges on the underlying dynamics.
	\section{Classifying two-to-two scattering processes}
	\label{sec.scattering}
\begin{figure}[t!]
		\centering
		    \includegraphics[width=.35\textwidth]{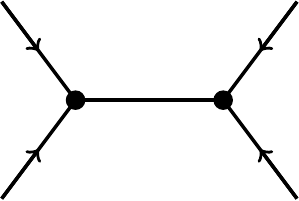} \qquad \quad
		    \includegraphics[width=.35\textwidth]{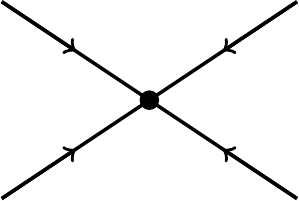}
		\caption{\label{fig:generaldiagrams} Tree-level Feynman diagrams contributing to a two-to-two scattering process. They correspond to particle-mediated interactions (left) and four-point interactions (right).  Each external line can either symbolize a scalar or a photon. The effective vertices are denoted by a black circle. We adopt the convention that the momentum of external particles point into the diagram.}
	\end{figure}

	%
	Following up on the classification given in the previous section, our interest is in physical observables related to particle scattering. The scattering amplitudes describing the most general two-to-two scattering process with external scalars and photons in a flat  spacetime are readily obtained from the effective action using standard Feynman diagram techniques. Since we are working with the dressed propagators and vertices derived from the effective action, all quantum corrections are already captured by tree-level diagrams. A process then receives contributions from Yukawa-type interactions as well as matter self-interactions related to the effective four-point vertices. The corresponding diagrams are schematically depicted in \autoref{fig:generaldiagrams}. 
	In this section, we summarize the most general amplitudes related to gravity-mediated scalar scattering (\autoref{sec.scalarscattering}), gravity-mediated photon scattering (\autoref{sec.photonscattering}), and mixed amplitudes involving both external scalars and photons (\autoref{sec.mixedscattering}). In the case where the scalar is taken very massive, these amplitudes can be used to describe the bending of light by a massive gravitational source. Moreover, a formalism to extract corrections to the Newtonian potential based on amplitude computations has been proposed in \cite{Donoghue:1993eb,Donoghue:1994dn,Bjerrum-Bohr:2014zsa,Bjerrum-Bohr:2017dxw}. Apart from the generalization considered in \autoref{sect.beyondflat}, we will work in a flat spacetime with the Minkowski metric given by $\eta_{\mu\nu} = \mathrm{diag}(+1,-1,-1,-1)$, such that the graviton $h_{\mu\nu}$ is defined by $g_{\mu\nu} = \eta_{\mu\nu} + h_{\mu\nu}$. This allows us to work with momentum-space techniques.
	
	\subsection{Kinematics of two-to-two scattering processes}
	\label{sect.3.1}
	As indicated by the arrows in \autoref{fig:generaldiagrams}, we adopt the convention that all external momenta flow into the diagram. We label these momenta by $p^\mu_i$ with $i=1,2$ referring to the ``incoming'' and $i=3,4$ to the ``outgoing'' particles. The on-shell condition for photons is
	\begin{equation}\label{eq.photononshell}
	p^2 = 0 \, . 
\end{equation}
	Scalars are taken to be massive,
	\begin{equation}\label{eq.scalaronshell}
	p^2 = m^2 \, . 
	\end{equation}
	
	It is convenient to parameterize the amplitude in terms of the Mandelstam variables
	\begin{equation}\label{eq.mandelstam}
	 \mans = (p_1 + p_2)^2 \, , \qquad \mant = (p_1 + p_3)^2 \, , \qquad \manu = (p_1 + p_4)^2 \, ,
	\end{equation}
	subject to 
	\begin{equation}\label{eq.mandelstamrelation}
	\mans + \mant + \manu = p_1^2 + p_2^2 + p_3^2 + p_4^2 \, . 
	\end{equation}
	Here $\mans$ gives the square of the center-of-mass energy (invariant mass) and $\mant$ is the square of the four-momentum transfer. 
	
	Denoting the spatial momenta by bold quantities, the external momenta can be parameterized by
	\begin{equation}\label{eq.momentapara}
	\begin{aligned}
	 p_{1\mu} = (\sqrt{m^2 + {\bf p}^2}, {\bf p}) \, , \qquad & 
	 	p_{2\mu} = (\sqrt{m^2 + {\bf p}^2}, -{\bf p}) \, , \\
	 p_{3\mu} = (-\sqrt{m^2 + {\bf q}^2}, {\bf q}) \, , \qquad & 
	 	p_{4\mu} = (-\sqrt{m^2 + {\bf q}^2}, -{\bf q}) \, ,
	\end{aligned}
	\end{equation}
	with photons coming with $m^2 = 0$. Defining the scattering angle $\theta$ by
	\begin{equation}
	{\bf p} \cdot {\bf q} = \sqrt{{\bf p}^2 \, {\bf q}^2} \cos\theta \, , 
	\end{equation}
	allows to express the Mandelstam variables $\mant$ and $\manu$ in terms of the center-of-mass energy $\mans$ and the scattering angle. For instance, for ingoing scalars with mass $m_1$ and outgoing scalars with mass $m_2$ one has
	\begin{equation}\label{eq.mandelanglerelation}
	\begin{aligned}
	 \mant = & \, - \left( \frac{\mans}{2} - m_1^2 - m_2^2 + \frac{1}{2} \sqrt{\left(\mans-4m_1^2\right)\left(\mans-4m_2^2\right)} \cos\theta \right) \, , \\	
	 \manu = & \, - \left( \frac{\mans}{2} - m_1^2 - m_2^2 - \frac{1}{2} \sqrt{\left(\mans-4m_1^2\right)\left(\mans-4m_2^2\right)} \cos\theta \right) \, .
	\end{aligned}
\end{equation}
 Finally, we introduce polarization vectors for photons. The ingoing polarization vectors are taken in the $(y-z)$-plane,
\begin{equation}\label{eq.polin}
e^{{\rm in}+}_\mu = \frac{1}{\sqrt{2}} \, (0,0,1,-i) \, , 
\end{equation}
and the outgoing polarization vectors are
\begin{equation}\label{eq.polout}
e^{{\rm out}+}_\mu =  \frac{1}{\sqrt{2}} \, (0,-\sin\theta,\cos\theta,-i) \, .
\end{equation}
The parameterizations in eqs.\ \eqref{eq.momentapara}, \eqref{eq.polin}, and \eqref{eq.polout} allow to work out all scalar products between momentum four-vectors and polarization tensors in terms of either the scattering angle and spatial momenta or, equivalently, using the Mandelstam variables \cite{Knorr:2022lzn}.

	\subsection{Amplitudes from the Quantum effective action}
	\label{sec.amplitudeencyclopidia}
	In this section, we will list the scattering amplitudes of two-to-two particle processes parameterized by the effective action presented in \autoref{sec.ea}. We will begin with scalar scattering in \autoref{sec.scalarscattering}. We then continue with photon scattering in \autoref{sec.photonscattering}, and conclude our encyclopaedia in \autoref{sec.mixedscattering} with scalar-photon scattering.
	
	\subsubsection{Gravity-mediated scalar scattering}
	\label{sec.scalarscattering}
	We begin our list of scattering processes with scalar scattering.
	The total amplitude for the process $\phi\phi \to \phi \phi$ is given by \cite{Draper:2020knh}
	\begin{equation}\label{eq.scalaramplitude}
			\mathcal{A}^{\phi}
		=
				\mathcal{A}^{\phi}_{\mans}
			+	\mathcal{A}^{\phi}_{\mant}
			+	\mathcal{A}^{\phi}_{\manu}
			+	\mathcal{A}^{\phi}_{4}
		\,\text{.}
	\end{equation}
	This corresponds to the scattering amplitude of the $\mans$, $\mant$ and $\manu$ channels, as well as the four-point diagram.
	
	The $\mans$-channel amplitude for scalar-to-scalar scattering reads
	\begin{equation}\label{eq.scalarschannel}
		\begin{aligned}
			\mathcal A_\mans^{\phi} 
		&= 
			\frac{4\pi \GN}{3} \bigg[
			 \left( (\mans + 2m^2) (1 + \mans f_{\text{Ric}\phi\phi}) - 12 \mans f_{R\phi\phi} \right)^2 G_{RR}(\mans) 
		\\&\hspace{9em} 
			 - \left(1+ \mans f_{\text{Ric}\phi\phi} \right)^2 G_{CC}(\mans) \Big\{ \mant^2 - 4 \mant \manu + \manu^2 \Big\}
		\bigg] \, .
		\end{aligned}
	\end{equation}
	Here we have evaluated the form factors $f_{\text{Ric}\phi\phi}$ and $f_{R\phi\phi}$ at
	\begin{equation}
		\begin{aligned}
			f_{\text{Ric}\phi\phi}	=	f_{\text{Ric}\phi\phi}(\mans,m^2,m^2)
		&\quad\text{and\quad}&
			f_{R\phi\phi}=			f_{R\phi\phi}(\mans,m^2,m^2)
		\,\text{.}\end{aligned}
	\end{equation}
	Furthermore, we have introduced the functions
	\begin{equation}\begin{aligned}\label{eq.propagatorfunctions}
			G_{RR}(x)
		=
			\frac{1}{x\, \left(1+x \, f_{RR}(x)\right)}
		\,\text{,}&\quad&
			G_{CC}(x)
		=
			\frac{1}{x\, \left(1+x\, f_{CC}(x)\right)}
		\,\text{,}
	\end{aligned}\end{equation}
	related to the graviton propagator. We obtain the $\mant$ and $\manu$ channels from \eqref{eq.scalarschannel} by crossing symmetry, interchanging $\mans \leftrightarrow \mant$ and $\mans \leftrightarrow \manu$, respectively.
	
	The four-point diagram is given by
	\begin{equation}\label{eq.scalar4point}
		\mathcal A_4^{\phi}
		=
		f_{\phi^4}\left(
		\frac{\mans-2m^2}{2},
		\frac{\mant-2m^2}{2},
		\frac{\manu-2m^2}{2},
		\frac{\manu-2m^2}{2},
		\frac{\mant-2m^2}{2},
		\frac{\mans-2m^2}{2}
		\right) \,.
	\end{equation}
	This concludes the description of the scalar scattering amplitude.
	\subsubsection{Gravity-mediated photon scattering}
	\label{sec.photonscattering}
	We will now consider four-photon scattering. Again, we have particle-mediated and four-point contributions to the scattering amplitude. 
	For the particle-mediated diagram, the exchanged particle is either a photon or a graviton. Computing the vertices arising from the action \eqref{eq.GammaA3}, and setting the two external photon legs on-shell shows that  the three-photon vertices vanish.
	Therefore, the only contribution comes from a graviton-exchanged diagram.
	
	The full amplitude for this process is given by \cite{Knorr:2022lzn}
	\begin{equation}
		\label{eq.photonamplitude}
		\mathcal{A}^{\gamma} = \mathcal{A}^{\gamma}_{\mans} + \mathcal{A}^{\gamma}_{\mant} + \mathcal{A}^{\gamma}_{\manu} + \mathcal{A}^{\gamma}_{4}
		\,\text{.}
	\end{equation}
	The amplitudes can be organized by their helicity configurations. We have the following classes:
	\begin{equation}\label{eq.helicityclasses}
		\begin{aligned}
			&\text{I:}	\qquad&	\mathcal{A}^{+--+} &= \mathcal{A}^{-++-} \,\text{,}
			\\
			&\text{II:}	\qquad&	\mathcal{A}^{++++} &= \mathcal{A}^{----} \,\text{,}
			\\
			&\text{III:}	\qquad&	\mathcal{A}^{++--}	&=	\mathcal{A}^{--++} \,\text{,} 
			\\
			&\text{IV:}	\qquad&	\mathcal{A}^{+-+-}	&=	\mathcal{A}^{-+-+} \,\text{,}
			\\
			&\text{V:}	\qquad&	\mathcal{A}^{+++-} &= \mathcal{A}^{++-+} = \mathcal{A}^{+-++} = \mathcal{A}^{-+++} =
			\\&& \mathcal{A}^{---+} &= \mathcal{A}^{--+-} = \mathcal{A}^{-+--} = \mathcal{A}^{+---}	\,\text{.}
		\end{aligned} 
	\end{equation}
	The $\mant$-channel contributions to these expressions read
	\begin{align}
		\text{I:}\quad&
		\mathcal{A}^{\gamma}_{\mant}=2 \pi G_N \mans^2 \big(-2 + \mant^2 \, f_{D^2\text{Ric}FF} - 2 \mant \, f_{\text{Ric}FF}  - 4 \mant \, f_{\text{Rm}FF}\big)^2	G_{CC}(\mant)
	 \,\text{,}\label{eq.photontchannel1}
	\\
		\text{II}=\text{IV:}\quad&
		\mathcal{A}^{\gamma}_{\mant}=\begin{aligned}[t]
			&	\frac{\pi }{3} \GN \mant^2	\big(-\mant \, f_{D^2\text{Ric}FF} + 4 f_{\text{Rm}FF} - f_{FF}'(0)\big)^2
			\\
			&\hspace{13em}\times
			 (\mans^2 - 4 \mans \manu + \manu^2) G_{CC}(\mant)
			\\	
			&  \hspace{-1em}- \frac{\pi }{3} \GN \mant^4\big(6 \mant \, f_{D^2RFF} + \mant \, f_{D^2\text{Ric}FF} - 24 f_{RFF} 
			\\&\hspace{6em}
			-6  f_{\text{Ric}FF}-4  f_{\text{Rm}FF} - 2  f_{FF}'(0)\big)^2 G_{RR}(\mant) \,\text{,}   
		\end{aligned}\label{eq.photontchannel2}
		\\
		\text{III:}\quad&
		\mathcal{A}^{\gamma}_{\mant}=2 \pi \GN \manu^2 \big(-2 + \mant^2 \, f_{D^2\text{Ric}FF} - 2 \mant \, f_{\text{Ric}FF}  - 4 \mant \, f_{\text{Rm}FF}\big)^2 G_{CC}(\mant) \,\text{,}\label{eq.photontchannel3}
		\\
		\text{V:}\quad&
		\mathcal{A}^{\gamma}_{\mant}=\begin{aligned}[t]	2 \pi 
			&	\GN \mans \manu \big(-\mant^2 f_{D^2\text{Ric}FF} + 4 \mant \, f_{\text{Rm}FF} - \mant \, f_{FF}'(0)  \big) 
			\\
			&	\qquad \times  \big(- 2 + \mant^2 \, f_{D^2\text{Ric}FF} - 2 \mant \, f_{\text{Ric}FF}  - 4 \mant \, f_{\text{Rm}FF}\big) G_{CC}(\mant) \,\text{.} 
		\end{aligned}\label{eq.photontchannel5}
	\end{align}
	Here the arguments of the form factors are suppressed in the following way:
	\begin{equation}
		\label{eq.onshellff}
		\begin{aligned}
			f_{RFF} &= f_{RFF}(\mant, 0, 0)	
			\, \text{,}	&\qquad
			f_{\text{Ric}FF} &= f_{\text{Ric}FF}(\mant, 0, 0)	
			\, \text{,}	&\qquad\\
			f_{D^2RFF} &= f_{D^2RFF}(\mant, 0, 0)	
			\, \text{,}	&\qquad
			f_{D^2\text{Ric}FF} &= f_{D^2\text{Ric}FF}(\mant, 0, 0)	
			\,\text{,}&\qquad \\
			f_{\text{Rm}FF} &= f_{\text{Rm}FF}(0, 0)	
			\,\text{.}
		\end{aligned}
	\end{equation}
	We now present the four-point amplitudes. Since these expressions are rather lengthy, we summarize them in \autoref{tab.photon4point}.
	\begin{table}[t!]\centering
		\renewcommand{\arraystretch}{1.5}
			\resizebox{\textwidth}{!}{
				\begin{tabular}{c|ccccc}
				&	I																																																				&	II																																																						&	III																																																					&	IV																																																																									&	V			\\
				&	$+--+$																																																		&	$++++$																																																				&	$++--$																																																				&	$+-+-$																																																																							&	$+++-$	\\\hline
				
				$f_{F^2F^2}$	&	$4\mans^2[f(\mans,\mant,\manu)+f(\mans,\manu,\mant)]$																																			&	$4\mant^2[f(\mant,\mans,\manu)+f(\mant,\manu,\mans)]$																																					&	$4\manu^2[f(\manu,\mans,\mant)+f(\manu,\mant,\mans)]$																																					&	$\text{I}+\text{II}+\text{III}$																																																															&	$0$										\\
				$f_{F^4}$		&	$\mans^2[\text{perm.}]$																																													&	$\mant^2[\text{perm.}]$																																															&	$\manu^2[\text{perm.}]$																																															&	$\begin{aligned}-2\{\mans\manu&[f(\mans,\mant,\manu)+f(\manu,\mant,\mans)]] \\ +\mans\mant&[f(\mans,\manu,\mant)+f(\mant,\manu,\mans)]\\+\mant\manu&[f(\mant,\mans,\manu)+f(\manu,\mans,\mant)]\}\end{aligned}$				&	$0$										\\
				$f_{FFDFDF_1}$			&	$\begin{aligned}\tfrac{1}{2}\mans^2\{	t&[f(\mans,\mant,\manu)-f(\mant,\mans,\manu)]	\\+ u&[f(\mans,\manu,\mant)-f(\manu,\mans,\mant)]	\}\end{aligned}$	&	$\begin{aligned}\tfrac{1}{2}\mant^2\{	\mans&[f(\mant,\mans,\manu)-f(\mans,\mant,\manu)]	\\+ \manu&[f(\mant,\manu,\mans)-f(\manu,\mant,\mans)]	\}\end{aligned}$	&	$\begin{aligned}\tfrac{1}{2}\manu^2\{	\mans&[f(\manu,\mans,\mant)-f(\mans,\manu,\mant)]	\\+ \mant&[f(\manu,\mant,\mans)-f(\mant,\manu,\mans)]	\}\end{aligned}$	&	$\frac{1}{2}\mans\mant\manu[\text{perm.}]$																																																											&	$0$										\\
				$f_{FFDFDF_2}$			&	$0$																																																			&	$0$																																																					&	$0$																																																					&	$0$																																																																								&	$0$										\\
				$f_{FFDFDF_3}$			&	$0$																																																			&	$0$																																																					&	$0$																																																					&	$\mans\mant\manu[\text{perm.}]$																																																															&	$\frac{1}{4}\mans\mant\manu[\text{perm.}]$	\\
				$f_{FFDFDF_4}$			&	$\mans^2[\mant f(\mans,\manu,\mant)+\manu f(\mans,\mant,\manu)]$																															&	$\mant^2[\mans f(\mant,\mans,\manu)+\manu f(\mant,\manu,\mans)]$																																	&	$\manu^2[\mans f(\manu,\mant,\mans)+\mant f(\manu,\mans,\mant)]$																																	&	$\text{I}+\text{II}+\text{III}$																																																															&	$-\frac{1}{2}\mans\mant\manu[\text{perm.}]$	\\
				$f_{FFD^2FD^2F}$	&	$-\frac{1}{2}\mans^2\mant\manu[f(\mans,\mant,\manu)+f(\mans,\manu,\mant)]$																											&	$-\frac{1}{2}\mant^2\mans\manu[f(\mant,\mans,\manu)+f(\mant,\manu,\mans)]$																														&	$-\frac{1}{2}\manu^2\mans\mant[f(\manu,\mans,\mant)+f(\manu,\mant,\mans)]$																														&	$\text{I}+\text{II}+\text{III}$																																																															&	$\frac{1}{2} \cdot \text{IV}$		\\
				
		\end{tabular} 
	}
		\caption{Different contributions of the four-photon vertex to $\mathcal{A}^{\gamma}_4$. The function $f$ should be read as $f(a,b,c)= f_I\left(\frac{a}{2},\frac{b}{2},\frac{c}{2},\frac{c}{2},\frac{b}{2},\frac{a}{2}\right)$. By $[\text{perm.}]$, we denote the sum of $f(\mans,\mant,\manu)$ and the five permutations of its arguments. The $16$ polarisation configurations can be obtained by interchanging $( + \leftrightarrow -)$ following the scheme in \eqref{eq.helicityclasses}. The total amplitude for each polarisation configuration is obtained by summing the contribution from each form factor in the respective column. (From \cite{Knorr:2022lzn}.)}\label{tab.photon4point}
	\end{table}

	\subsubsection{Mixed amplitudes}
	\label{sec.mixedscattering}
	Finally, we consider the process $\gamma\phi \to \gamma \phi$. Since the two particles in the initial and final states are distinguishable, only the $\mant$-channel will give a contribution to the exchange diagrams. The total amplitude is thus given by
	\begin{equation}\label{eq.scalarphotonamplitude}
			\mathcal{A}^{\gamma\phi}
		=
			\mathcal{A}^{\gamma\phi}_{\mant}
		+	\mathcal{A}^{\gamma\phi}_{4}
		\,\text{.}
	\end{equation}
	The $\mant$-channel amplitude is given by
	\begin{align}
		\mathcal{A}^{++}_{\mant}
		&= 	\begin{aligned}[t]
			&	-\frac{2 \pi \GN }{3} 	
				\left(1 - \mant f_{\text{Ric}D\phi D\phi} \right)	\left(\mant f_{D^2\text{Ric}FF} - 4 f_{\text{Rm}FF} + f_{FF}'(0) \right)
			\\&\hspace{12em}\times
			\left(\mans^2 - 4\mans \manu + \manu^2 + 2m^4	\right) \mant G_{CC}(\mant) \\
			&+	\frac{2 \pi \GN }{3} \mant^{2} 
				\left(\mant + 2m^2 + 2\mant \left(\mant - m^2 \right) f_{\text{Ric}D\phi D\phi}- 12 \mant  f_{R\phi \phi} \right)
			G_{RR}(\mant)\\
			& \hspace{-1em}\times \big(6 \mant f_{D^2RFF} + \mant f_{D^2\text{Ric}FF} - 24 f_{RFF} -6 f_{\text{Ric}FF}-4 f_{\text{Rm}FF} - 2 f_{FF}'(0) \big) \,\text{,}
		\end{aligned}\label{eq.scalarphoton++} \\
		\mathcal{A}^{+-}_{\mant}
		&=		\begin{aligned}[t]&
			4 \pi \GN 
			\left(1 - \mant f_{\text{Ric}D\phi D\phi} \right) \left(-2 + \mant^2 \, f_{D^2\text{Ric}FF} - 2 \mant f_{\text{Ric}FF}  - 4 \mant \, f_{\text{Rm}FF} \right)
		\\& \hspace{18em}\times
		\left(\mans \manu - m^4\right) G_{CC}(\mant) \,\text{.}
		\end{aligned}\label{eq.scalarphoton+-}
	\end{align}
	Here we have suppressed the arguments of the form factors as in \eqref{eq.onshellff}, and in addition
	\begin{equation}\begin{aligned}\label{eq.onshellff2}
		f_{R\phi  \phi} = f_{R\phi \phi}(\mant, m^2, m^2) 
	\,\text{,} &\qquad&
		f_{\text{Ric}D\phi D \phi} = f_{\text{Ric}D\phi D\phi}(\mant, m^2, m^2) 
	\,\text{.}
	\end{aligned}\end{equation}
	We continue with the four-point amplitude. This is given by
	\begin{align}
		\mathcal{A}_4^{++}
		&=		\begin{aligned}[t]\frac{1}{4}	\mant \bigg[&
			8	f_{FF\phi\phi}(\mant,\mans,\manu)
			-	2 \mans	f_{FF D\phi D\phi}(\mant,\mans,\manu)
			+	2m^2 f_{FFD^2\phi \phi}(\mant,\mans,\manu)
			\\&\hspace{8em}
			-	(\mans\manu - m^4)	f_{FFD^2\phi D^2\phi}(\mant,\mans,\manu)
			\bigg]
			+	(\mans \leftrightarrow \manu)
			\,\text{,}
		\end{aligned}
		\\
		\mathcal{A}_4^{+-}
		&=
		\frac{1}{4}( \mans\manu-m^4)	\bigg[
		2	f_{FFD\phi D\phi}(\mant,\mans,\manu)
		-	2	f_{FFD^2\phi\phi}(\mant,\mans,\manu)
		\\&\hspace{14.5em}
		-	\mant	f_{FFD^2\phi D^2\phi}(\mant,\mans,\manu)
		\bigg]
		+	(\mans \leftrightarrow \manu)
		\,\text{.}
	\end{align}
	Here, we suppressed half of the arguments of the form factors following the prescription
	\begin{equation}
		f_{PQRS}(a,b,c) = f_{PQRS}\left(\frac{a}{2},\frac{b-m^2}{2},\frac{c-m^2}{2},\frac{c-m^2}{2},\frac{b-m^2}{2},\frac{a}{2}-m^2\right)
		\, .
	\end{equation}
	From the amplitudes $\mathcal{A}^{++}$ and $\mathcal{A}^{+-}$, the two remaining helicity configurations can then be obtained by parity transformations. This gives
	\begin{align}
		\mathcal{A}^{++} = \mathcal{A}^{--}	\,\text{,}
		\qquad
		\mathcal{A}^{+-} = \mathcal{A}^{-+}	\,\text{.}
	\end{align}
	Finally, the scattering amplitude of the  process $\phi\phi \to \gamma\gamma$ can be obtained by using crossing symmetry. Since the polarization channels do not change, we obtain the scattering amplitude from \eqref{eq.scalarphotonamplitude} by exchanging $\mant \leftrightarrow \mans$. This concludes our listing of scattering amplitudes.
	
	\subsection{Going beyond flat spacetime}
	\label{sect.beyondflat}
	A natural generalization of the amplitude formalism developed above is to go beyond a flat spacetime. Promoting scattering amplitudes to a curved  spacetime allows to study aspects of the non-linear coupling of quantum fields to gravity.
	In \autoref{sec.curvedQFTreview}, we discuss some of the challenges that arise in the formulation of QFT on a curved spacetime. Many of these problems are already present in the simplest setting where spacetime exhibits a constant scalar curvature. We present some recent developments in the description of scattering amplitudes in such a spacetime in \autoref{sec.dSamplitudes}.
	
	\subsubsection{Challenges in QFT on a curved spacetime}\label{sec.curvedQFTreview}
	
	The generalization of QFT from Minkowski spacetime to curved spacetime comes with a number of major challenges. Here we highlight two: the absence of non-interacting asymptotically flat boundaries and the lack of spacetime symmetries.
	
	Generic curved spacetimes do not possess an asymptotically flat boundary. Solutions to the wave equation interact at any time with the gravitational field and are not propagating freely asymptotically: it can be shown that even one-particle states can decay \cite{Polyakov:2007mm,Anderson:2013ila,Anderson:2017hts,Markkanen:2016aes}.
	As a consequence, it is problematic to describe a scattering process as well-defined particles entering the scattering region and interacting for a limited time. This is also the main obstruction to constructing an $S$-matrix 	\cite{Witten:2001kn,Bousso:2004tv,Marolf:2012kh,Mandal:2019bdu,Giddings2013TheGS}.
	
	Another complication in the formulation of QFT in a curved spacetime is related to a potential lack of symmetries. In Minkowski spacetime, there is a unique distinguished vacuum state that is Poincar\'e invariant. Particles defined with respect to this vacuum are characterized by conserved energy-momentum charges. In general, such a state does not exist in a curved setting and particles cannot be classified according to conserved quantities. Also, at a computational level, amplitudes in Minkowski spacetime are most conveniently computed using momentum-space techniques. Due to the non-commutativity of covariant derivatives, there exists no momentum-space in curved spacetime. This is another major obstacle in the concrete calculation of curved-spacetime scattering amplitudes.
	
	\subsubsection{Scattering amplitudes in de~Sitter spacetime}\label{sec.dSamplitudes}
	
	De~Sitter spacetime serves as a benchmark for the construction of scattering amplitudes in curved spacetime. This maximally-symmetric space is characterized by constant positive scalar curvature $R=12H^2$, where $H$ is Hubble's constant, while all other curvature tensors vanish.
	
	QFT in de~Sitter spacetime already exhibits many of the problems present in the construction of QFT in a general curved spacetime, and in particular the challenges posed above. However, due to the relatively simple curvature structure, properties such as the (non-)uniqueness of the vacuum can be studied explicitly.
	
	The simple curvature structure also lends itself for the form factor framework. Since any derivative of the spacetime curvature vanishes, we can classify interactions in an expansion about a constantly-curved reference point instead of about Minkowski spacetime. The form factors are now characterized by an additional curvature parameter. For example, the $R\phi\phi$-form factor generalizes to
	\begin{equation}\label{eq.dSformfactor}
		\int \dd^4x\sqrt{-g} \, f^{\text{dS}}_{R\phi\phi}(R,\dalembertian_1,\dalembertian_2,\dalembertian_3) R\phi\phi
		\,\text{.}
	\end{equation}
	This action contains the interactions that contribute to the flat three-point vertex, since setting $f_{R\phi\phi}(\dalembertian_1,\dalembertian_2,\dalembertian_3) = f_{R\phi\phi}^{\text{dS}}(0,\dalembertian_1,\dalembertian_2,\dalembertian_3)$ gives the $R\phi\phi$-term contained in $\Gamma_{h\phi^2}$ in \eqref{eq.Gammahphi2}. One has to keep in mind, however, that the form factor in \eqref{eq.dSformfactor} overlaps with the generalized kinetic term. The extraction of vertices and propagators in de~Sitter spacetime is therefore a subtle question.
	
	At this point, a remark about the cosmological constant $\Lambda$ is in order. When previously working in a flat spacetime, we have explicitly set $\Lambda$ to zero. This ensured that the Minkowski spacetime is a solution to the gravitational equations of motion. In order to render de~Sitter spacetime an on-shell solution, the inclusion of a positive $\Lambda$ is required.
	
	Having discussed the problems related to the classification of the interactions that contribute to a de~Sitter spacetime amplitude, the next task would be to compute the associated tree-level Feynman diagrams. Conceptually, this is a straightforward generalization of the procedure in flat spacetime: vertices and propagators are obtained from $\Gamma$ by taking functional derivatives with respect to the fields, while the Feynman diagram is constructed from the contraction of the vertices and propagators.
	
	In order to bring the scattering amplitude in a manifestly gauge-invariant and on-shell form, we have to carefully take into account the commutator of covariant derivatives with form factors. Techniques for handling such commutators were developed in the context of affine gravity in \cite{Knorr:2020bjm} and adapted to metric gravity in \cite{Ferrero:2021lhd}.
	
	Using these techniques, we can compute a scattering amplitude in terms of differential operators. In the case of a conformally-coupled  massless scalar field in Einstein-Hilbert gravity, the tree-level scattering amplitude functional of the graviton-mediated process $\phi\phi \to \phi\phi$ in de~Sitter spacetime reads
	\begin{equation}\label{eq.dSamplitude}
		\mathcal{A}^\phi_{\text{dS}} = 16\pi \GN	\int \dd^4 x\sqrt{-g}\, T_{\alpha\beta}[\phi_1,\phi_2] \mathcal{G}(\dalembertian) T^{\alpha\beta}[\phi_3,\phi_4]
		\,\text{.}
	\end{equation}
	Here the propagator $\mathcal{G}(\dalembertian)$ is given by
	\begin{equation}\label{eq.dSpropagator}
		\mathcal{G}(\dalembertian) = \left(\dalembertian+2H^2\right)^{-1}
		\,\text{,}
	\end{equation}
	while the on-shell vertex is captured by the tensor
	\begin{equation}\label{eq.dSvertex}
		\begin{aligned}
			T_{\mu\nu}[\phi_1,\phi_2] 
			&=
				(\CD_{(\mu}\phi_1)(\CD_{\nu)}\phi_2)
				-	\frac{1}{4}g_{\mu\nu}	(\CD_\gamma\phi_1)(\CD^\gamma\phi_2)
				\\&\hspace{3cm}
				-	\frac{1}{6}\left[\CD_\mu\CD_\nu + \frac{1}{4}g_{\mu\nu}\dalembertian\right]\phi_1\phi_2
				\,\text{.}
		\end{aligned}
	\end{equation}
	Comparing this expression to the scattering amplitude in flat spacetime, we see that covariant derivatives play the role of generalized momenta. We notice, however, that these are fully non-commuting operators due to the finite spacetime curvature.
	
	Due to the differential operator nature of \eqref{eq.dSpropagator}, computing the number given by \eqref{eq.dSamplitude} is less straightforward than in flat spacetime. In \cite{Ferrero:2021lhd}, the scattering amplitude of massive scalars in Einstein-Hilbert gravity was computed in the adiabatic limit, where the scalar mass $m$ is much larger than the Hubble constant $H$. This is the de~Sitter spacetime analog of taking the non-relativistic limit. By Fourier-transforming the resulting amplitude, one obtains the Newtonian potential $V$. For small separations $r$, this reproduces the standard $-1/r$ potential, in accordance with flat spacetime. For separations $r \lesssim H$, one obtains curvature corrections that can be interpreted as a repulsive force. This complies with the picture of de~Sitter spacetime as an exponentially expanding FLRW universe. For super-Hubble radii $r>H$, the potential is identically zero, thereby making manifest that fields separated by the de~Sitter horizon are not in causal contact.

	\section{Asymptotically safe scattering amplitudes}
	\label{sec.tanh}
	The amplitudes constructed in \autoref{sec.scattering} constitute the most general result for scattering in a flat spacetime compatible with a relativistic QFT. The goal of this section is to explore Asymptotic Safety within this general framework. Our discussion focuses on the gravity-mediated scattering of scalar particles in a flat spacetime, building on the results of \autoref{sec.scalarscattering}. We start by summarizing the properties of \emph{asymptotically safe amplitudes} in \autoref{sect.51} before providing instructive examples in \autoref{sect.52}. Some general remarks are added in \autoref{sect.54}. 
	
	Our exposition is limited to scattering processes with four external fields. This implies that phenomena related to IR divergences in the amplitudes, whose cure is expected to result from the resummation of diagrams including an arbitrary number of soft external gravitons \cite{Weinberg:1965nx}, is beyond the scope of the exposition. Similarly, discussing the back-reaction of the scattered particles onto the geometry is also beyond the scope of this chapter, as incorporating this effect in a flat spacetime may also require adding additional gravitons. While this renders the exposition incomplete in some aspects, it nevertheless highlights the importance of from factors for the gravitational Asymptotic Safety program.
	\subsection{Definition of an asymptotically safe amplitude}
	\label{sect.51}

 The idea that gravity could be asymptotically safe appeared in the seminal work by Weinberg \cite{Weinberg:1980gg}. This essay provided the following characterization of the Asymptotic Safety mechanism: ``\emph{A theory is said to be asymptotically safe if the essential coupling parameters approach a fixed point as the momentum scale of their renormalization point goes to infinity. This condition is introduced [$\ldots$] as a means of avoiding unphysical singularities at very high energy.}'' This statement of intent can actually be found in many works exploring the viability of this scenario by studying the Wilsonian RG flow for gravity and gravity-matter systems.
 
 The scattering amplitudes constructed in the previous section constitute prototypical entities which should be free from unphysical divergences. Formulating Asymptotic Safety in terms of the form factor framework then allows to make this statement of intent precise in an operational way. The notion of essential couplings translates to the combination of form factors that appear within an on-shell amplitude. The fact that the amplitudes are gauge-invariant illustrates that not all combinations of form factors are essential in this sense: since the gauge-fixing does not appear in the amplitude, it is inessential according to this definition. The concept of the momentum scale of an essential coupling is captured by the generalized momentum dependence of the form factors. 
 Notably, the form factor framework can capture more theories than those with a ``running'' coupling depending on a single momentum scale.
 They capture the general momentum dependence of the $n$-point vertices on all independent combinations of the external momenta.
 
 Asymptotic Safety manifests itself as constraints on the momentum dependence of the combinations building \emph{essential form factors}: they should remain bounded as the center-of-mass energy goes to infinity. This constraint applies to \emph{all} amplitudes that can be constructed from a given field content. This implies, in particular, that all amplitudes are bounded in the Jin-Martin sense \cite{Jin:1964zza}, satisfying
 \begin{equation}\label{jin-martin}
 \lim_{|\mans| \rightarrow \infty} \mans^{-2} \mathcal{A}(\mans,\mant) = 0 \, .
 \end{equation}
 The necessary cancellations require a delicate interplay between the form factors making up the propagators and vertices. It is this point where the RG fixed points underlying Asymptotic Safety come into play. If the high-energy behavior of a theory is controlled by such a UV fixed point, quantum scale invariance may precisely provide the necessary relations for canceling the unphysical divergences.
 
	\subsection{Realizing asymptotically safe amplitudes via form factors}
 	\label{sect.52}
For concreteness, let us consider the gravity-mediated scattering of two massless scalar particles $\phi, \chi$ with a standard kinetic term,
\begin{equation}
f_{\phi\phi} (\dalembertian) = f_{\chi\chi}(\dalembertian) = \dalembertian \,\text{.}
\end{equation}
Throughout the discussion in this subsection, we assume that the scalars are minimally coupled to gravity, so that all form factors associated with non-minimal gravity-matter vertices vanish,
\begin{equation}\label{ff.approx}\begin{aligned}
		 f_{R\phi\phi} = f_{\text{Ric}\phi\phi} = f_{R\chi\chi} = f_{\text{Ric}\chi\chi} = 0
		\,\text{.}
	\end{aligned}	
\end{equation}

Let us first focus on the two-to-two scattering process $\phi\phi \rightarrow \chi \chi$.
In this case, the scattering amplitude $\mathcal{A}_\mans(\mans,\theta)$ is encoded by a single $\mans$-channel diagram whose topology is given by the left diagram of \autoref{fig:generaldiagrams}:
\begin{equation}\label{eq.5.1}
\mathcal{A}_\mans(\mans,\theta) = \frac{4\pi \GN}{3} \, \mans^2 \, \left[ G_0(\mans) - P_2(\cos\theta) \, G_2(\mans) \right] \, . 
\end{equation}
Here $\theta$ is the scattering angle in the center-of-mass frame, and $P_j(x)$ are the standard Legendre polynomials of order $j$ with $P_2(x) \equiv (3x^2-1)/2$. The scalar parts of the spin-zero and spin-two propagators including the form factor contributions \eqref{eq.Gammah2} are
\begin{equation}\label{eq.5.2}
\begin{aligned}
G_0^{-1}(p^2) = (p^2 + \imath \epsilon) \left(1+p^2 \, f_{RR}(p^2) \right) \, , \\
G_2^{-1}(p^2) = (p^2 + \imath \epsilon) \left(1+p^2 f_{CC}(p^2) \right) \, .  
\end{aligned}
\end{equation}
The massless pole has been equipped with the standard ``$\imath\epsilon$'' prescription for a Feynman propagator. The form factor contributions may have similar terms in order to give the correct prescription for poles and branch cuts. These terms are left implicit.

The $\mans$-channel scattering process is conveniently analyzed using the partial wave decomposition
\begin{equation}\label{eq.5.3}
a_j(\mans) \equiv \frac{1}{32\pi} \int_{-1}^1 \dd \cos\theta \, P_j(\cos\theta) \, \mathcal{A}_\mans(\mans,\cos\theta) \, . 
\end{equation}
The partial wave amplitudes encode the dependence of the spin $j$ part of the amplitude on the center-of-mass energy $\mans$. The process \eqref{eq.5.1} is then described by two non-zero partial wave amplitudes that capture the dependence of the amplitude on the spin-zero and spin-two part of the graviton propagator \eqref{eq.5.2},
\begin{equation}\label{eq.5.4}
a_0(\mans) = \frac{\GN}{12} \, \mans^2 \, G_0(\mans) \, , \qquad a_2(\mans) = - \frac{\GN}{60} \, \mans^2 \, G_2(\mans)\, . 
\end{equation}

The analysis of the process $\phi\chi \rightarrow \phi\chi$ is more complicated. In this case, the amplitude also receives contributions from the $\mant$- and $\manu$-channel. The pole in the $\mant$-channel diagram leads to a divergence of the amplitude in the forward scattering limit. As a consequence, this process cannot be analyzed by a partial wave decomposition. However, this divergence is an IR effect and related to the massless nature of the graviton. Hence, its status is different from the UV divergences in the focus of our discussion.

\subsubsection{General Relativity} 
Let us return to the case  $\phi\phi \rightarrow \chi \chi$. The $\mans$-channel amplitude for this process in General Relativity (GR) is
\begin{equation}\label{schannel-gr}
\mathcal{A}_\mans^{\phi\phi \rightarrow \chi\chi,{\rm GR}} = \frac{4 \pi \GN}{3} \left( \mans - \frac{\mant^2 - 4 \mant \manu +  \manu^2}{\mans} \right) \, . 
\end{equation}
The partial wave amplitudes describing tree-level scattering  are obtained from \eqref{eq.5.4} by setting the form factors to zero. In this way, one recovers the well-known result
\begin{equation}\label{eq.5.5}
a_0^{\rm GR}(\mans) = \frac{\GN}{12} \, \mans \, , \qquad a_2^{\rm GR}(\mans) = - \frac{\GN}{60} \, \mans \, . 
\end{equation}
As their characteristic feature, these partial wave amplitudes grow without bound when the center-of-mass energy approaches infinity. \autoref{fig:Stelle} illustrates the analytic properties of $a_2(\mans)$. 

The growth of the amplitude can already be deduced based on dimensional analysis. The partial wave amplitudes must be dimensionless. The topology of the tree-level diagram furthermore implies that the result must be proportional to $\GN$. In order to compensate the negative mass dimension of Newton's constant, the amplitude has to come with positive powers of the center-of-mass energy. This argument readily extends to loop corrections appearing at higher orders in perturbation theory. These terms have to come with higher powers of the dimensionless quantity $\GN \mans$, aggravating the divergence. While this argument does not exclude that going beyond perturbation theory and performing a resummation of the series could lead to a well-defined UV behavior of the amplitude, it shows that treating graviton-mediated scattering within GR leads to unphysical divergences at high energy, at least at the perturbative level. 

For the subsequent discussion, it is useful to extend this analysis to the process $\phi\chi \rightarrow \phi\chi$. This amplitude is obtained from \eqref{schannel-gr} by crossing symmetry, $\mans \leftrightarrow \mant$, and reads
\begin{equation}\label{tchannelpole}
	\mathcal{A}_\mant^{\phi\chi \rightarrow \phi\chi,\text{GR}} = 8\pi \GN \frac{\mans \manu}{\mant}	=	-	8\pi \GN \frac{\mans(\mans+\mant)}{\mant}
	\,\text{.}
\end{equation}
For forward scattering, $\mans \to \infty $ while keeping $\mant$ fixed, this amplitude diverges quadratically. This exhibits two additional difficulties in analyzing gravity-mediated scattering amplitudes within GR. Firstly, the divergence in the forward scattering limit makes it difficult to apply the optical theorem. Secondly, the amplitude is not bounded in the Jin-Martin sense \cite{Jin:1964zza}, \ie{}, it violates the constraint \eqref{jin-martin}.

The general discussion of \autoref{sec.amplitudeencyclopidia} then suggests that form factors can provide the crucial contributions that tame these divergences and render the amplitude finite. It is instructive to understand the underlying mechanisms based on two specific sets of form factors: the ones obtained in the framework of quadratic gravity \cite{Stelle:1976gc} and the tanh-model investigated in \cite{Draper:2020bop,Knorr:2021iwv}.

\subsubsection{Quadratic Gravity} 
Before moving towards the discussion of asymptotically safe amplitudes, it is instructive to take a short detour and review the prototype of the graviton-mediated asymptotically free scattering amplitudes realized in the framework of quadratic gravity (QG). In the language of form factors, \eqref{eq.Gammah2}, the bare action of this theory is obtained by taking $f_{CC}(x)$ and $f_{RR}(x)$ to be constant,
\begin{equation}\label{eq.ffqgbare}
f_{CC}^{QG,{\rm bare}} = - \frac{1}{(m_2^{\rm bare})^2} \, , \qquad 
f_{RR}^{QG,{\rm bare}} = - \frac{1}{(m_0^{\rm bare})^2} \, .
\end{equation}  
Here $m_0^{\rm bare}$ and $m_2^{\rm bare}$ are the bare masses of the additional massive degrees of freedom accompanying the massless graviton familiar from GR.
\begin{figure}[t]
	\includegraphics[width = 0.5\textwidth]{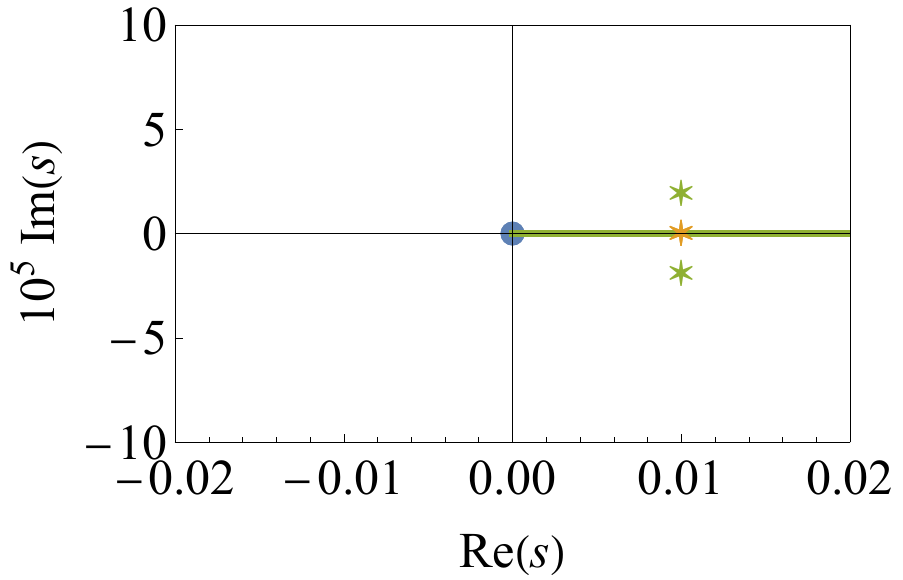}%
	\includegraphics[width = 0.5\textwidth]{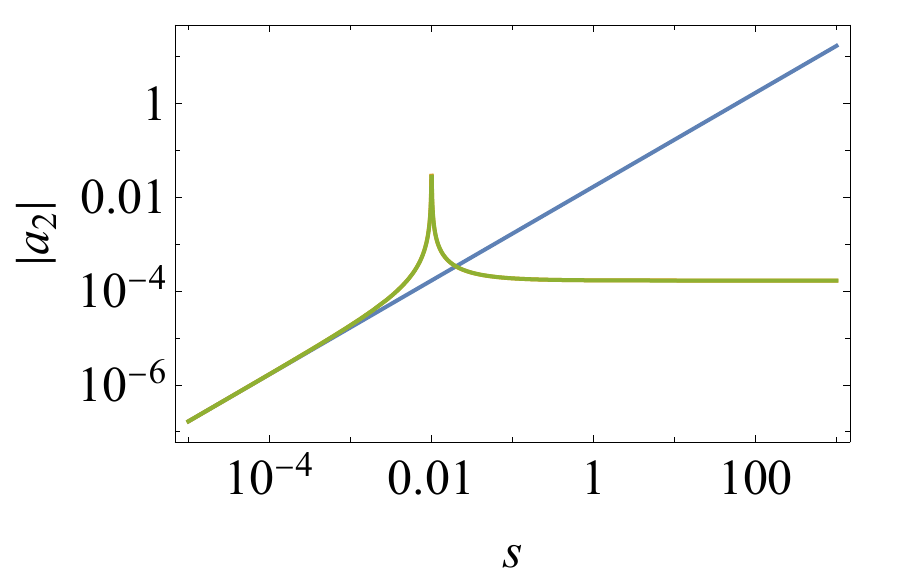}
	%
	%
	\caption{Illustration of the graviton-mediated scattering process $\phi\phi \rightarrow \chi\chi$ described within quadratic gravity. The analytic properties of the spin-two propagator are displayed in the left panel. At tree-level, $G_2(p^2)$ exhibits a massless pole (blue dot) and a ghost pole (orange star). Including the one-loop correction, the latter splits into a pair of complex-conjugate poles (green stars) and a branch cut (green line). The partial wave amplitude $a_2(\mans)$ is shown in the right panel. For center-of-mass energies below the ghost pole, the amplitude follows the one found in GR (blue line). At $\mans \approx (m_2)^2$, the amplitude for quadratic gravity exhibits a resonance. At energies larger than the ghost pole, the tree-level amplitude approaches a constant, while the inclusion of the one-loop correction to the propagator (green line) leads to a logarithmic fall-off. The latter reflects that quadratic gravity is asymptotically free. 
		The illustration is obtained for $(m_2)^2 = 0.01$, $\epsilon = 10^{-7}$, and $N_{\rm eff} = 23/6$.	
	For the selected parameter values, the tree-level and the one-loop amplitudes are virtually indistinguishable, showing that one ghost pole (orange) and the combination of two complex-conjugate poles and a branch cut (green) lead to very similar amplitudes.	}
	\label{fig:Stelle}
\end{figure}
 The presence of these additional degrees of freedom is readily seen by plugging \eqref{eq.ffqgbare} into \eqref{eq.5.2} and investigating the scalar part of the tree-level propagators. For instance, the spin-two propagator takes the suggestive form
\begin{equation}\label{eq.qgproptree}
G_2^{\rm tree}(p^2) = \frac{1}{p^2} - \frac{1}{p^2 - (m_2^{\rm bare})^2} \, . 
\end{equation}
Thus the massive degree of freedom is a ghost, coming with a ``wrong sign'' in the propagator.\footnote{Despite being a negative norm state, it has been argued that the ghost does not lead to a violation of unitarity, since its short lifetime implies that it is not an asymptotic state. In the modern interpretation due to Donoghue et.\ al.\ \cite{Donoghue:2019ecz} and Anselmi et.\ al.\ \cite{Anselmi:2018tmf,Anselmi:2018bra}, this additional degree of freedom has the status of a Merlin mode, moving backward in time, thereby violating causality at Planckian time scales. It is tempting to speculate that such an effect is the manifestation of quantum fluctuations in the light cone structure of spacetime at the level of the effective action \cite{Donoghue:2021meq}.} The analytic structure of the propagator and the resulting partial wave amplitude $a_2$ are illustrated in orange in \autoref{fig:Stelle}. This shows that the ghost mode tames the growth of the amplitude at center-of-mass energies exceeding the pole mass. As a consequence, $\lim_{\mans\rightarrow\infty} a_2(\mans)$ is finite. The ghost thereby acts like a Pauli-Villars regulator, curing the UV divergence in the amplitude.

Upon including the one-loop self-energy corrections to \eqref{eq.qgproptree}, the scalar part of the propagator receives logarithmic corrections \cite{Donoghue:2018lmc},
\begin{equation}\label{eq.qgprop1loop}
\left( G_2^{\rm 1-loop}(p^2) \right)^{-1} =  \, p^2 + i \epsilon - \frac{p^4}{m_2^2} - \frac{\GN}{20 \pi}  \, p^4 \, N_{\rm eff} \, \log\left(\frac{-p^2 - i \epsilon}{\mu^2} \right)  \, . 
\end{equation}
Here $\mu^2$ is an energy scale making the argument of the logarithm dimensionless, and $N_{\rm eff} = N_V + N_F/4 + N_S/6 + 21/6$ counts the number of light degrees of freedom comprising $N_V$ vectors, $N_F$ fermions, and $N_S$ scalars. In our example with two massless scalar fields $N_{\rm eff} = 23/6$.

The one-loop correction affects the analytic structure of the propagator, converting the ghost pole into a pair of complex conjugate poles and a branch cut. This is illustrated in green in the left diagram of \autoref{fig:Stelle}. Moreover, as shown in the right diagram of \autoref{fig:Stelle}, the amplitude now decreases logarithmically for $\mans \gg m_2^2$. This reflects that quadratic gravity is asymptotically free. This conclusion holds modulo the caveat that the form factors related to non-minimal gravity-matter interactions \eqref{ff.approx}, that have not been included in the analysis, do not modify the fall-off of the amplitudes in the UV.

While the last property indicates that quadratic gravity is not asymptotically safe in the sense envisioned by Weinberg, one can nevertheless draw some important lessons from this example. First, \eqref{eq.5.2} allows to readily recast the quantum corrections to the propagator in a one-loop form factor,
\begin{equation}\label{eq.ff.oneloop}
f_{CC}^{\rm QG,1-loop}(x) = - \frac{1}{m_2^2} - \frac{\GN}{20 \pi} \, N_{\rm eff} \, \log\left(\frac{-x - i \epsilon}{\mu^2} \right) \, , 
\end{equation}
where the $i\epsilon$ selects the correct branch of the logarithm. On this basis, we expect that the effective action contains form factors with a non-trivial momentum dependence and a specific analytic structure in terms of branch cuts and poles. They play a crucial role in understanding the stability properties of resonances and graviton bound states that determine the high-energy behavior of the theory. Owed to the long-range nature of gravity, the form factors will also contain non-local terms.

Second, the discussion illustrates the link between form factors and results obtained from standard perturbative quantization techniques. Denoting the dimensionless coupling multiplying the Weyl-squared term by $1/\xi^2$, the one-loop contribution \eqref{eq.qgprop1loop} can be understood as the logarithmic running of this coupling with respect to the physical momentum scale.\footnote{Generally, this running should be discriminated from the scale dependence of couplings arising from integrating out quantum fluctuations shell-by-shell in a Wilsonian RG approach. For example, Newton's coupling has a non-trivial scale dependence with respect to the Wilsonian coarse graining scale $k$ \cite{Reuter:1996cp,Reuter:2001ag}. At the same time, the structure of the effective action dictates that $\GN{}$ cannot be promoted to a form factor, and therefore must be constant at the level of $\Gamma$. See \cite{Bonanno:2020bil} for a detailed discussion.} At the level of the effective action, this running is captured by a momentum-dependent form factor. In the case of $1/\xi^2$, this is given by $f_{CC}(\Delta)$.

\subsubsection{The tanh-model} 

We now discuss a model that realizes physical asymptotic safety in the manner originally proposed by Weinberg \cite{Weinberg:1980gg}. We stress at this point that this model does not arise from a first-principle calculation, but was constructed to illustrate how the quantum effective action formalism can efficiently include concepts such as Asymptotic Safety, unitarity and causality. The model was constructed and discussed in \cite{Draper:2020bop}.

The model again considers the graviton-mediated scattering of two minimally coupled scalar fields $\phi$ and $\chi$ with the form factors associated with the non-minimal gravity-matter couplings set to zero, cf.\ \eqref{ff.approx}. The non-trivial form factors are present in the gravitational sector; we choose
\begin{equation}\label{eq.tanhformfactors}\begin{aligned}
		f_{RR}(\dalembertian) = c_{RR}	\GN	\tanh\left(c_{RR}\GN\dalembertian\right)
		\,\text{,}&\qquad&
		f_{CC}(\dalembertian) = c_{CC}	\GN	\tanh\left(c_{CC}\GN\dalembertian\right)
		\,\text.
\end{aligned}\end{equation}
Here $c_{RR},c_{CC} >0$ are numerical parameters that control the scale where the non-trivial form factors become significant. Finally, by specifying the scalar four-point amplitude, we fix the four-point form factor $f_{\phi^2\chi^2}$. This gives the four-point amplitude
\begin{equation}
	\mathcal{A}_4^{\phi\chi} = f_{\phi^2\chi^2}\left(\frac{\mans}{2},\frac{\mant}{2},\frac{\manu}{2},\frac{\manu}{2},\frac{\mant}{2},\frac{\mans}{2}\right)	+	\text{sym}
	\,\text{,}
\end{equation}
where ``sym'' indicates symmetrization of the arguments of $f_{\phi^2\chi^2}$ due to the functional variation of the action. The four-point amplitude is effectively parameterized by a function of three arguments, that we choose to be
\begin{equation}\label{eq.tanh4point}\begin{aligned}
		\mathcal{A}_4^{\phi\phi\to\chi\chi} = g(\mans|\mant,\manu)
		&\qquad&
		g(a|x,y) = 4\pi \GN G_{CC}(a)	(x^2+y^2)	f^{\text{int}}(a^2+x^2+y^2)
		\,\text{.}
\end{aligned}\end{equation}
Here $G_{CC}$, defined in \eqref{eq.propagatorfunctions}, is fixed by the graviton form factor $f_{CC}$. The function $f^{\text{int}}$ interpolates between zero and one, and is chosen to be
\begin{equation}
	f^{\text{int}}(x) = \frac{c_t \GN^2 x \tanh\left(c_t \GN^2	x\right)}{1+c_t \GN^2 x \tanh\left(c_t \GN^2 x\right)}
	\,\text{.}
\end{equation}
Here $c_t>0$ is a dimensionless parameter that controls the scale where the four-point form factor becomes significant.
Plugging in these form factors into \eqref{eq.scalaramplitude}, we obtain the scattering amplitude for the $\phi\phi\to \chi\chi$ process.

\begin{figure}[t]
	\includegraphics[width = 0.5\textwidth]{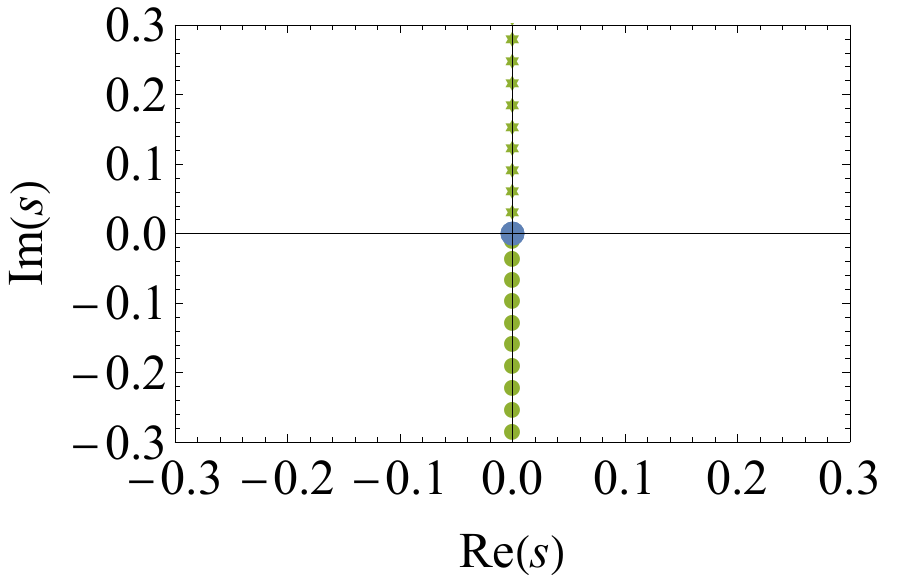}%
	\includegraphics[width = 0.5\textwidth]{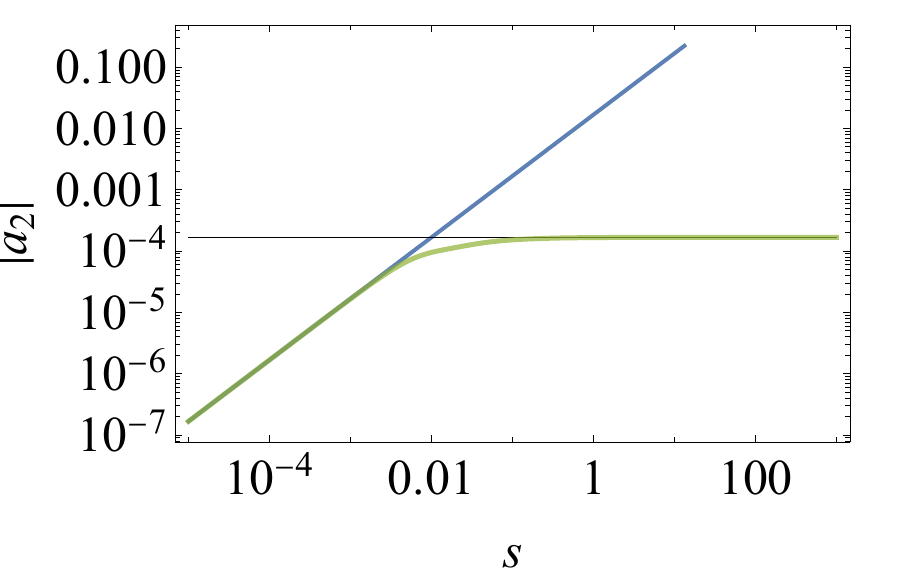}
	\caption{Left panel: The analytic structure of the partial wave amplitude $a_2(\mans)$ for the $\tanh$-model. Green dots and stars mark poles at imaginary squared momentum with positive and negative residues, respectively. The blue dot gives the massless graviton pole from GR. Right panel: The partial wave amplitude $|a_2(\mans)|$ for the $\tanh$-model is shown as the green line while the result of GR has been added as the blue line for reference. The black horizontal line shows the asymptotic value $\lim_{\mans\to\infty}|a_2(\mans)| = 1/6000$. The model parameters are $\GN=1$ and $c_{CC} = 100$. At this point, we did not include the scalar self-interaction.}
	\label{fig:tanhpwa}
\end{figure}

We analyze the properties of the $\mans$-channel amplitude via the partial wave decomposition. We find the same partial wave amplitudes as in \eqref{eq.5.4}.
Both partial wave amplitudes share the same basic behavior; for brevity, we will therefore only discuss the partial wave amplitude $a_2$. It is plotted in \autoref{fig:tanhpwa}.
From this plot, we see that for small center-of-mass energies $\mans$, the partial wave amplitude $a_2$ does not differ from the GR amplitude. However, for large energies, rather than growing linearly with $\mans$, the amplitude $a_2$ becomes constant, reaching the asymptotic values
\begin{equation}\begin{aligned}
		\lim_{\mans\to\infty} a_0(\mans) = \frac{1}{12}	\frac{1}{c_{RR}}
		\,\text{,}&\qquad&
		\lim_{\mans\to\infty} a_2(\mans) = -	\frac{1}{60}	\frac{1}{c_{CC}}
		\,\text{.}
\end{aligned}\end{equation}
This is in agreement with the expectation from unitarity that the partial wave amplitudes are bounded by one, $|a_j(\mans)| \leq 1$.

In order to claim that a setting is asymptotically safe, it does not suffice that just one amplitude remains bounded: all scattering amplitudes that can be constructed from the field content under consideration must be free from unphysical divergences. For the gravity-mediated scalar scattering, this implies that also the amplitude for the process $\phi\chi \rightarrow \phi\chi$ must be finite. The consequences of this requirement have been analyzed in detail in \cite{Draper:2020bop} and we summarize the central insights obtained from this case.
We start from $\mathcal{A}_\mans^{\phi\chi}$ and use the crossing symmetry $\mans \leftrightarrow \mant$. Taking the GR limit, the resulting amplitude is given by \eqref{tchannelpole}.
For forward scattering, $\mans \to \infty $ while keeping $\mant$ fixed, this amplitude diverges quadratically in $\mans$. Inserting the form factors \eqref{eq.tanhformfactors} does not improve this behavior.

\begin{figure}[t!]\centering
		\includegraphics[width=.49\textwidth]{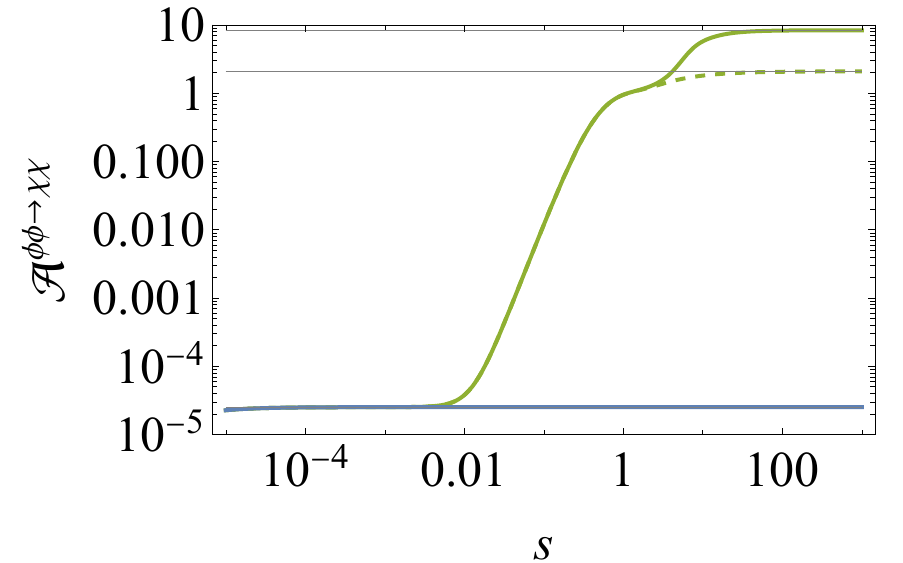}\hfill
		\includegraphics[width=.49\textwidth]{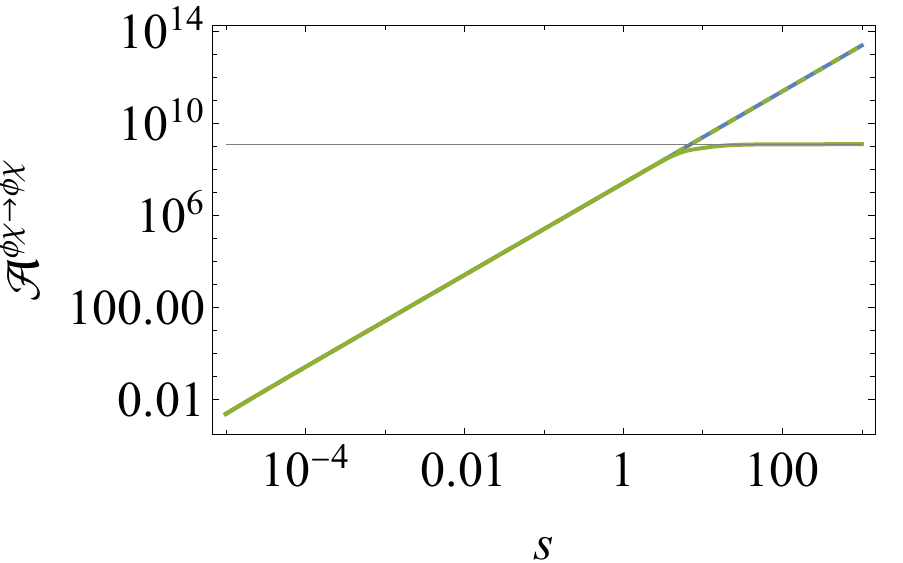}
	\caption{%
		The scattering amplitudes  $\mathcal{A}^{\phi\phi\to\chi\chi}$ (left panel) and $\mathcal{A}^{\phi\chi\to\phi\chi}$ (right panel) in the forward scattering limit for GR (blue) and the $\tanh$-model (green). The solid line denotes the amplitude with self-interaction, while the dashed line depicts the $\mans$-channel diagram only. In these plots, $\mant = -10^{-6}\,\GN^{-1}$ is held fixed, with parameters $c_{RR} = 1$, $c_{CC}=2$ and $c_t=10^{-2}$. In both panels, asymptotes are denoted by thin gray lines.
	}
	\label{fig.tanh4point}
\end{figure}

At this point, the four-point interaction becomes crucial. The four-point amplitude is readily derived from \eqref{eq.tanh4point} by using crossing symmetry; we obtain
\begin{equation}
	\mathcal{A}_4^{\phi\chi\to\phi\chi} = g(\mant|\mans,\manu)
	\,\text{.}
\end{equation}
The total scattering amplitude $\mathcal{A}^{\phi\chi\to\phi\chi} = \mathcal{A}_\mant^{\phi\chi} + \mathcal{A}_4^{\phi\chi\to\phi\chi}$ is shown in the right panel of \autoref{fig.tanh4point}. Here, we see that the total amplitude becomes scale-free in the forward scattering limit. Therefore, the interplay between the graviton propagator and the four-point interaction ensures the boundedness of the total scattering amplitude. In addition, this does not spoil the finiteness properties of the amplitude in the $\phi\phi \to \chi\chi$ process (left panel in \autoref{fig.tanh4point}).

At first sight, the necessary tuning of the matter self-interactions based on properties of the graviton propagator may appear artificial. From the perspective of Asymptotic Safety, such relations are not unexpected though: the UV behavior of an asymptotically safe theory is governed by an interacting fixed point of the Wilsonian RG. The enhanced symmetry associated with the fixed point should then manifest itself in such a way that \emph{all} scattering amplitudes remain bounded. Thus it is not inconceivable that the seemingly ad hoc relations between form factors in the high-energy limit have a natural explanation in terms of deeper symmetry principles underlying the construction.
	
Notably, this symmetry principle applies to the UV limit of the amplitudes only. In particular, we expect that there is a substantial freedom in how asymptotically safe amplitudes interpolate between their UV and IR parts. In the $\tanh$-model this freedom is reflected, for instance, in the choice of the interpolation function $f^{\rm int}(x)$. In the language of the RG, this freedom is reflected by selecting a specific RG trajectory flowing away from the fixed point. The free parameters characterizing this choice (and parameterizing the UV critical surface of the fixed point) will also appear at the level of physical observables, \eg{}, determining the class of amplitudes being asymptotically safe with respect to this specific fixed point.

Finally, we observe that the $\tanh$-model exhibits very peculiar properties with respect to causality. Focusing on the scalar part of the spin-two propagator, we find that its analytic structure encodes the contribution of the massless graviton supplemented by an infinite tower of unstable resonances coming with a mass and decay time set by Planckian scales. While the graviton sets a definite arrow of time, the resonances are distributed very symmetrically with respect to their causality properties. For each fixed resonance, there is exactly one positive and one negative energy mode which propagates forward and backward in time \cite{Platania:2022gtt}. Hence the model is expected to violate microcausality on Planckian time scales. Whether this reflects a generic feature of asymptotically safe amplitudes is currently unknown.

\subsection{Anomalous dimensions within the form factor framework}
\label{sect.54}
Let us finally make the connection between form factors and the anomalous dimension of the graviton propagator. Following \cite{Platania:2022gtt}, we define the anomalous dimension of a propagator $G_i(p^2)$ in the UV by
\begin{equation}\label{eq.anomdim}
\eta_i^\infty \equiv \lim_{p^2 \rightarrow \infty} \left( - p \frac{\p}{\p p} \, \ln\left( \, p^{-2} \, G_i^{-1}(p^2) \,  \right) \right) \, . 
\end{equation} 
The relation between the form factors and the scalar parts of the propagators  is given in \eqref{eq.5.2} (with $\epsilon =0$). Now suppose that, for large momenta $p^2$, the form factor under consideration follows the power law scaling
\begin{equation}\label{eq.ffscaling}
f_i(p^2) \simeq (p^2)^a \, , \qquad a \in \mathbb{R} \,  .
\end{equation}
It is then straightforward to evaluate \eqref{eq.anomdim}, yielding
\begin{equation}
\eta_i^\infty = \left\{
\begin{array}{cl}
	-2(1+a) \qquad & a > -1 \, , \\
	0 & a < -1 \, . 
\end{array} \right.
\end{equation}
We observe that, generically, there is a cap for the anomalous dimension. For $a < -1$ the scaling of the two-point function in the UV is no longer dominated by the form factor \eqref{eq.ffscaling} but fixed by the Einstein-Hilbert contribution to the scalar part of the propagator. The fact that this transition among the leading behavior happens at $a=-1$ follows from the observation that the form factor contribution $f_i(p^2)$ is accompanied by a prefactor $(p^2)^2$ arising from the two spacetime curvature terms evaluated in the flat spacetime. Thus the Einstein-Hilbert term and the form factor exhibit the same UV scaling if $a=-1$. Hence, if the form factors are dictating the high energy behavior of the propagator, the natural value for the UV anomalous dimensions obtained in the context of form factors \emph{is negative} when compared to the $p^2$-scaling of the propagator arising from the Einstein-Hilbert action.

However, this does not entail that form factors cannot accommodate positive values $\eta_i^\infty$. Choosing 
\begin{equation}\label{eq.fftuned}
f_i(p^2) = - (p^2)^{-1} + (p^2)^{\tilde a} \,  ( 1+\mbox{subleading}) \, , 
\end{equation}  
leads to 
\begin{equation}
\eta_i^\infty = -2 (1+ \tilde{a}) \, , \qquad \tilde{a} \in \mathbb{R} \, ,
\end{equation}
which is positive for $\tilde{a} < -1$. This requires the precise cancellation between non-local form factor terms and the Einstein-Hilbert contribution: the first term in \eqref{eq.fftuned} has been precisely constructed in such a way that it eliminates the contribution of the latter in $G_i(p^2)$.

In principle, it is also possible to construct asymptotically safe amplitudes based on graviton propagators exhibiting a UV enhancement (manifested by $\eta_i^\infty > 0$). In this case, the growth of the graviton propagator must be compensated by a suitable decay of the gravity-matter vertices similar to the cancellations induced by \eqref{eq.fftuned}. Thus the interplay between the different building blocks for such amplitudes must be even more intricate than in the case where $\eta_i^\infty < 0$.

\section{Connection to low-energy effective field theory}\label{sec.eft}
Asymptotic Safety has the ambition to provide a description of gravity and matter valid on all momentum scales. The discussion of asymptotically safe amplitudes demonstrated that form factors are an essential element in the construction. Formulating Asymptotic Safety in terms of the effective action $\Gamma$ also provides a direct link to low-energy effective field theory. This link allows to test the construction based on positivity bounds expected to hold for the low-energy theory in order to admit a UV completion.\footnote{This discussion has a natural connection with the ``swampland program'' investigated in the context of string theory \cite{Palti:2019pca}. At this point it should be stressed \cite{Basile:2021krr} that UV completions based on string theory and Asymptotic Safety may entail different definitions of the swampland since the space of effective field theories embeddable into string theory and admitting a UV completion through a Wilsonian RG fixed point may be different.}

Conceptually, the link between the effective action $\Gamma$ of an asymptotically safe theory and its low-energy effective action is quite simple. Starting from $\Gamma$, one introduces a UV scale $\Lambda_{\rm UV}$ and express all couplings and momenta in terms of dimensionless quantities constructed from this scale. Subsequently, one expands $\Gamma$ in (inverse powers of) the UV scale and truncates the expansion at a fixed order. In this way, one obtains the low-energy effective field theory capturing the physics at scales below $\Lambda_{\text{UV}}$. Technically, this expansion turns the form factor framework into a derivative expansion, restricting the retained interactions to the ones with a low mass dimension only. Provided that the form factor classification is carried out to sufficiently high order, the resulting expansion is consistent in the sense that it captures all interactions that appear in the effective field theory, potentially including quantum corrections.

The drawback of working in the derivative expansion is that it may lead to spurious poles in propagators and Ostrogradski-type instabilities \cite{Becker:2017tcx} that may be absent in the full theory. Thus identifying the correct analytic structure of a propagator in this setting is far from trivial \cite{Platania:2020knd} and extra care has to be taken when interpreting the result. On the positive side, the derivative expansion directly connects to positivity bounds derived for the low-energy effective field theory. Typically, these bounds are obtained on the basis that the theory should come with a high-energy completion that is Lorentz-invariant, local, and causal \cite{Pham:1985cr,Ananthanarayan:1994hf,Adams:2006sv}. In particular, the optical theorem, stating that the imaginary part of the forward limit of an amplitude with the same initial and final particle content $i$ is given by
\begin{equation}\label{eq.optical}
{\rm Im} \, \mathcal{A}_{i}(\mans,\mant)|_{\mant =0} = \frac{1}{2} \sum_f \int \dd\Pi_f \, |\mathcal{A}_{i \rightarrow f}|^2 \geq 0 \, .  
\end{equation}
Here $\mathcal{A}_{i \rightarrow f}$ is the amplitude for the process $i \rightarrow f$ with $f$ denoting all possible admissible intermediate states and $\int \dd\Pi_f$ is the integral over the corresponding phase space. In the absence of negative norm states, all contributions on the right-hand side are positive so that each intermediate state gives a positive contribution to the imaginary part of the forward scattering amplitude. While the optical theorem is very powerful for non-gravitational theories, it is difficult to extend it to gravitational theories, see \cite{Tokuda:2020mlf,Alberte:2020bdz} for a recent discussion. The reason is the $\mant$-channel pole, \eqref{tchannelpole}, whose residue grows faster than the Jin-Martin bound \eqref{jin-martin}. Moreover, the pole at $\mant=0$ and the associated branch point arising from graviton loops lead to severe complications when performing the analytic continuation from $\mant < 0$ to $\mant \ge 0$. This is essential in the derivation of positivity bounds.

Despite these complications, it is worthwhile to discuss a specific example of such bounds derived in the context of Quantum Electrodynamics (QED) minimally coupled to gravity. The corresponding action takes the form
\begin{equation}\label{eq.lagQED}
S_{\rm QED} = \int \dd^4x\sqrt{-g} \left[ - \frac{1}{16\pi \GN} R - \frac{1}{4} F_{\mu\nu} F^{\mu\nu} +
\bar{\psi} \left( i \slashed{\CD} + m \right) \psi + e A_\mu \bar{\psi} \gamma^\mu \psi \right] \, . 
\end{equation}
The matter sector constitutes the prototype of a perturbatively renormalizable QFT. However, adding the minimal coupling to gravity makes the theory perturbatively non-renormalizable. Thus, \eqref{eq.lagQED} could be interpreted as an effective field theory. There is also evidence that gravity coupled to the matter content of QED admits UV completions via the Asymptotic Safety mechanism \cite{Harst:2011zx,deBrito:2020dta,Gies:2020xuh}.\footnote{Note that several swampland conjectures also constrain the UV completion of QED coupled to gravity, see, \eg{}, \cite{Arkani-Hamed:2006emk, Palti:2019pca}.} Our subsequent discussion of the low-energy bounds proposed for this system follows the exposition \cite{Cheung:2014ega}.

At energies below the mass of the electron, one may obtain an effective field theory capturing the dynamics of the photon and the graviton by integrating out the ``heavy'' electron. This results in the Euler-Heisenberg action
\begin{equation}\label{eq.eulerheisenberg}
\begin{aligned}
S_{\text{Eul-Heis}} = \int\dd^4 x\sqrt{-g} & \Big[ 
- \frac{1}{16 \pi \GN} R - \frac{1}{4} F_{\mu\nu} F^{\mu\nu} + a_1 \left( F_{\mu\nu} F^{\mu\nu} \right)^2 + a_2 \left( F_{\mu\nu} \tilde{F}^{\mu\nu} \right)^2 \\
& \; + b_1 R \,  F_{\mu\nu} F^{\mu\nu} + b_2 R_{\mu\nu}  F^{\mu\lambda} F^{\nu}{}_\lambda + b_3 R_{\mu\nu\lambda\sigma} F^{\mu\nu} F^{\lambda\sigma} \\
& \; + c_1 R^2 + c_2 R_{\mu\nu} R^{\mu\nu} + c_3 R_{\mu\nu\lambda\sigma} R^{\mu\nu\lambda\sigma} + \cdots \Big] \, . 
\end{aligned}
\end{equation}
This structure follows from the classification in \autoref{sec.ea} by eliminating all interaction monomials containing derivatives acting on the field strength tensor and setting the form factors to constants. The Wilson coefficients $a_i$, $b_i$ and $c_i$ have been computed in \cite{Drummond:1979pp,Cheung:2014ega}, but are not essential for the argument.

The analysis of the physics consequences arising from \eqref{eq.eulerheisenberg} can be simplified by eliminating the Riemann-squared term through the Gauss-Bonnet identity, and converting the Riemann tensor in the gravity-matter coupling to the Weyl tensor. Subsequently, the tree-level Einstein equations,
\begin{equation}\label{eq.Einstein-tree}
R_{\mu\nu} - \frac{1}{2} g_{\mu\nu} R = 8 \pi \GN \left( \frac{1}{4} g_{\mu\nu} F_{\rho\sigma} F^{\rho\sigma} - F_{\mu\rho} F_\nu{}^\rho \right) \, , 
\end{equation}
are used to eliminated the remaining curvature-squared terms and non-minimal gravity-matter couplings. This is equivalent to removing the inessential couplings with an Einstein-Hilbert starting point, as discussed earlier in \autoref{sec:fieldredef}. This procedure results in
\begin{equation}\label{eq.Euler-Heisenberg-red}
\begin{aligned}
	S_{\text{Eul-Heis,2}} = \int \dd^4x\sqrt{-g} & \Big[ 
	- \frac{1}{16 \pi \GN} R - \frac{1}{4} F_{\mu\nu} F^{\mu\nu}  + a_1^\prime \left( F_{\mu\nu} F^{\mu\nu} \right)^2 \\ & \; + a_2^\prime \left( F_{\mu\nu} \tilde{F}^{\mu\nu} \right)^2
	 + b_3 \, C_{\mu\nu\lambda\sigma} F^{\mu\nu} F^{\lambda\sigma} + \cdots \Big] \, . 
\end{aligned}
\end{equation}
with the couplings being related by
\begin{equation}
\begin{aligned}
	a_1^\prime = & \, a_1 - 2 \pi \GN \, b_2 + 4 \pi \GN \, b_3 + \frac{1}{4} (8 \pi \GN)^2 \left(c_2 + 4 c_3 \right) \, , \\  
	a_2^\prime = & \, a_2  - 2 \pi \GN \, b_2 + 4 \pi \GN \, b_3 + \frac{1}{4} (8 \pi \GN)^2 \left(c_2 + 4 c_3 \right) \, . 
\end{aligned}
\end{equation}

The coefficients $a_1^\prime$ and $a_2^\prime$ can then be constrained based on general physics arguments. For instance, the absence of super-luminality in a certain electromagnetic configuration entails \cite{Cheung:2014ega}
\begin{equation}\label{eq.bound-superluminality}
a_1^\prime + a_2^\prime \ge 0 \, . 
\end{equation}
The same conclusion arises from analyzing the symmetrized scattering amplitude associated with light-by-light scattering,
\begin{equation}\label{eq.M-sym}
\mathcal{A}_{\rm sym} \equiv \mathcal{A}^{++--} + \mathcal{A}^{--++} + \mathcal{A}^{+--+} + \mathcal{A}^{-++-} \, . 
\end{equation}
Evaluating the general expression \eqref{eq.helicityclasses} for the specific couplings \eqref{eq.eulerheisenberg}, one obtains in the forward scattering limit
\begin{equation}\label{eq.forwardamp}
\lim_{\mant \rightarrow 0} \mathcal{A}_{\rm sym}(\mans,\mant)  = - \frac{32 \pi \GN \mans^2}{\mant} - 32  \pi \GN \mans + 32 (a_1^\prime + a_2^\prime) \mans^2 + \mathcal{O}(\mant) \, . 
\end{equation}
The first term is the $\mant$-channel pole, signaling the divergence of the amplitude in the forward limit. Requiring that the coefficient appearing in front of the $\mans^2$-term is positive again implies the bound \eqref{eq.bound-superluminality}.

Based on unitarity, one can even argue a more stringent bound, suggesting that both coefficients be positive individually,
\begin{equation}\label{eq.unitarity.bound}
a_1^\prime \ge 0 \, , \qquad a_2^\prime \ge 0 \, . 
\end{equation}
The underlying idea is that these effective couplings are generated by UV degrees of freedom that do not contain states of negative norm. The argument then develops along the following lines. To leading order, the coupling between the photon and these UV degrees of freedom is parameterized by
\begin{equation}\label{eq.coupling-parametric}
F^{\mu\nu} F^{\rho\sigma} \, \chi_{\mu\nu\rho\sigma} \, , \qquad 
F^{\mu\nu} \tilde{F}^{\rho\sigma} \, \psi_{\mu\nu\rho\sigma} \, . 
\end{equation}
The fields $\chi$ and $\psi$ are parity even and odd, respectively, and possess the same index symmetry as the Riemann tensor. Moreover, any couplings have been absorbed into these fields. In order to foster the analysis, they are decomposed into their trace and traceless parts, \ie{},
\begin{equation}\label{eq.chidecomp}
\chi_{\mu\nu\rho\sigma} = \chi^{(4)}_{\mu\nu\rho\sigma} + \frac{1}{4} \left( \eta_{\mu[\rho} \chi^{(2)}_{\sigma]\nu} - \eta_{\nu[\rho} \chi^{(2)}_{\sigma]\mu} \right) + \frac{1}{2} \chi^{(0)} \eta_{\mu[\rho} \eta_{\sigma]\nu} \, ,
\end{equation}
and similarly for $\psi$. The tensorial component fields $\chi^{(4)}_{\mu\nu\rho\sigma}$ and $\chi^{(2)}_{\mu\nu}$ are traceless by definition. One then stipulates that the propagators of the component fields admit a standard spectral representation
\begin{equation}\label{eq.spectral}
\begin{aligned}
\langle \chi^{(0)}(p) \, \chi^{(0)}(p^\prime) \rangle = & \, i \delta^4(p+p^\prime) \int_0^\infty \dd\mu^2 \, \frac{\rho^{(0)}(\mu^2)}{p^2-\mu^2+ i \epsilon} \, , \\
\langle \chi^{(2)}_{\mu\nu}(p) \, \chi^{(2)}_{\alpha\beta}(p^\prime) \rangle = & \, i \delta^4(p+p^\prime) \int_0^\infty \dd\mu^2 \, \frac{\rho^{(2)}(\mu^2)}{p^2-\mu^2+ i \epsilon} \,
\Pi_{\mu\nu\alpha\beta} \, , \\
\langle \chi^{(4)}_{\mu\nu\rho\sigma}(p) \, \chi^{(4)}_{\alpha\beta\gamma\delta}(p^\prime) \rangle = & \, i \delta^4(p+p^\prime) \int_0^\infty \dd\mu^2 \, \frac{\rho^{(4)}(\mu^2)}{p^2-\mu^2+ i \epsilon} \, \Pi_{\mu\nu\rho\sigma\alpha\beta\gamma\delta} \, .
\end{aligned}
\end{equation}
The spectral densities $\rho^{(i)}(\mu^2)$ are supposed to be positive definite and capture arbitrary collections of single- and multi-particle states. The $\Pi$ are constructed from the Minkowski metric $\eta_{\mu\nu}$ and four-momentum $p_\mu$, and carry the tensor structure of the correlators. A careful analysis of their properties reveals \cite{Cheung:2014ega} that $\Pi_{\mu\nu\rho\sigma\alpha\beta\gamma\delta}$ necessarily involves negative norm states. On this basis, it is excluded and the four-point interactions are assumed to be generated from $\chi^{(0)}$ and $\chi_{\mu\nu}^{(2)}$ only. Going to low energy, $p^2 \rightarrow 0$, one then finds spectral representations of the effective couplings $a_1^\prime$ and $a_2^\prime$. For instance
\begin{equation}\label{eq.specrep}
\begin{aligned}
F^{\mu\nu} F^{\rho\sigma} \, \chi_{\mu\nu\rho\sigma} \; \rightarrow \; & \frac{1}{12}\left(F_{\mu\nu} F^{\mu\nu} \right)^2 \int_0^\infty \dd\mu^2 \, \frac{6 \rho^{(0)}(\mu^2) + \rho^{(2)}(\mu^2)}{\mu^2 + i \epsilon}  \\ & \; 
+ \frac{1}{8} \left(F_{\mu\nu} \tilde{F}^{\mu\nu} \right)^2 \int_0^\infty \dd\mu^2 \, \frac{\rho^{(2)}(\mu^2)}{\mu^2 + i \epsilon} \, . 
\end{aligned}
\end{equation}
The bounds \eqref{eq.unitarity.bound} then arise from the positivity of the spectral integral. Notably, this is a weaker condition than demanding that $\rho^{(i)}(\mu^2)$ is a positive function. Therefore, the latter may not strictly be necessary to conclude that the couplings generated by the ultraviolet states are positive.

From the perspective of Asymptotic Safety, it is worthwhile to make the following observation. The argument leading to \eqref{eq.unitarity.bound} tacitly assumes that the corresponding operators are absent in the ultraviolet theory. Owed to the non-vanishing gravitational interactions present at the underlying RG fixed point, it is expected that these interactions are already present in the fixed point action \cite{Eichhorn:2012va} since they are compatible with the symmetries of the kinetic term of the photon and therefore evade the non-renormalization condition \cite{Laporte:2021kyp}. In this light, it is conceivable that a positive spectral function of the ultraviolet contributions rather reflects itself in a monotonicity property of the Wilsonian RG flow rather than a positivity bound of the form \eqref{eq.unitarity.bound} \cite{Becker:2014pea}.

Quite remarkably, the condition of causality also gives rise to constraints on the gravitational three-point vertex. Ref.\ \cite{Camanho:2014apa} analyzed these conditions for the case of a weakly coupled gravitational theory, concluding that certain higher-derivative corrections to the three-point vertex may lead to a time-advance in a high-energy scattering process that can overwhelm the classical Shapiro time-delay. In the context of quadratic gravity, these arguments have recently been refined in \cite{Edelstein:2021jyu}. Moreover, bounds from four-graviton scattering at one-loop level have recently been discussed in \cite{Bern:2021ppb}. It would be interesting to investigate to which extent form factors may alter these conclusions.

\section{Form factors from first principles}
\label{sec.hardcore}
So far, our discussion adopted a perspective that focused on the general form of a scattering amplitude in the presence of form factors. In general, any quantum theory of gravity is expected to make distinct predictions for the form factors. This fixes the properties of the amplitudes arising from the approach. In this section, we review the status of first principle computations of form factors within asymptotically safe quantum gravity. So far, two routes have been investigated to obtain non-perturbative form factors: functional RG methods on the one hand, and the reconstruction from Monte Carlo data on the other. These techniques and the corresponding results are described in \autoref{sect.6.1} and \autoref{sec.63}, respectively.

	\subsection{Form factors from the functional renormalization group}
	\label{sect.6.1}
	The first-principle construction of non-perturbative form factors based on functional RG methods predominantly employs the Wetterich equation \cite{Wetterich:1992yh,Morris:1993qb,Reuter:1996cp} in Euclidean signature spacetimes. This equation realizes the idea of the Wilsonian RG, integrating out quantum fluctuations shell-by-shell in momentum space. Concretely, it encodes the dependence of the effective average action $\Gamma_k$ on the coarse-graining scale $k$:
	\begin{equation}\label{FRGE}
	k \p_k \Gamma_k = \frac{1}{2} {\rm Tr}\left[ \left( \Gamma^{(2)}_k + R_k \right)^{-1} \, k \p_k R_k \right] \, . 
	\end{equation}
	Here $\Gamma_k^{(2)}$ denotes the second functional derivative of $\Gamma_k$ with respect to the fluctuation fields, $R_k$ is an IR regulator, equipping fluctuations with momenta $p^2 \lesssim k^2$ with a $k$-dependent mass term, and the trace contains an integral over loop-momenta as well as a sum over fields. The interplay between the regulator terms on the right-hand side ensures that the trace is UV finite, so that the change of $\Gamma_k$ is actually driven by integrating out fluctuations with momenta $p^2 \approx k^2$. In the context of gravity, the construction of \eqref{FRGE} requires the introduction of an undetermined background metric $\bar{g}_{\mu\nu}$, so that the spacetime metric $g_{\mu\nu}$ can be decomposed into its background part and arbitrary fluctuations $ h_{\mu\nu}$, \eg{}, through the linear split
	\begin{equation}\label{split}
	g_{\mu\nu} = \bar{g}_{\mu\nu} + h_{\mu\nu} \, . 
	\end{equation}
	The background then allows to construct a reference scale, discriminating UV from IR modes. More details on the Wetterich equation and its application to gravity can be found in the reviews \cite{Niedermaier:2006wt,Codello:2008vh,Reuter:2012id,Dupuis:2020fhh} as well as in the introductory chapter of this volume \cite{me1}.
	
	A standard strategy for extracting non-perturbative information from \eqref{FRGE} consists in making an ansatz for $\Gamma_k$ (called truncation), supplying a closure condition for the right-hand side, and subsequently computing the flow of the $k$-dependent quantities retained by the truncation. At the level of form factors, this procedure generically leads to complicated non-linear integro-differential equations that need to be solved numerically. At the technical level, it is useful to discriminate among computations determining form factors at the background level and the non-trivial momentum dependence in correlation functions built from the fluctuation fields. Owed to the presence of the gauge-fixing and regulator terms breaking the split symmetry \eqref{split}, the results do not necessarily agree \cite{Pawlowski:2020qer}. The former may then be reconstructed from the latter by solving additional split-Ward or Nielsen identities.
	\subsubsection{Background form factors}
	The computation of background form factors is closely linked to the classification of interaction monomials given in \autoref{sec.ea}. The truncation of $\Gamma_k$ is obtained by promoting the form factors appearing in $\Gamma$ to $k$-dependent functions. Using \eqref{FRGE}, one determines the flow of these functions with respect to the coarse graining scale, as well as their $k$-stationary forms at a RG fixed point. The evaluation of the trace can be done in a background-independent way employing (off-diagonal) heat-kernel techniques \cite{Decanini:2005gt,Benedetti:2010nr,Codello:2012kq}.
	
So far, these computations have mostly focused on the propagators and the gravitational form factors induced by free matter fields. The canonical example is the re-derivation of the Polyakov action in two dimensions \cite{Codello:2010mj}, which yields a non-local form factor contribution
	\begin{equation}
	 \Gamma^\text{Pol} = -\frac{1}{96\pi} \int \dd^2x \, \sqrt{g} \, R \, \frac{1}{\dalembertian} \, R \, ,
	\end{equation}
	for a single scalar field. For the physical case of four dimensions, the low-energy form factors have been systematically investigated in \cite{Satz:2010uu, Codello:2015oqa, Codello:2015mba, Ohta:2020bsc, Knorr:2021niv}, including some cosmological applications \cite{Codello:2015pga}.
	
	A first self-consistent computation of a gravitational form factor has been put forward in \cite{Bosma:2019aiu}, where $f_{CC}$ was computed within conformally reduced quantum gravity. Qualitatively, a very good fit for positive (Euclidean) momenta was given as
	\begin{equation}
	 f_{CC}(x) \approx f_\infty + \frac{\rho}{\frac{\rho}{\kappa} + x} \, ,
	\end{equation}
	for numerical parameters $f_\infty, \rho, \kappa$. Incidentally, this is very similar to what one finds for the propagator from RG improvement \cite{Bonanno:2000ep}.
	
	Following up on \cite{Becker:2017tcx}, a general toolbox to compute form factors from first principles was provided in \cite{Knorr:2019atm}. There, also $f_{\phi\phi}$ was computed, but almost no deviation from a free scalar field was found. A similar computation, but in a different gauge, was performed in \cite{Meibohm:2015twa}, where the deviations from a free propagator vanished identically for a massless scalar.
	\subsubsection{Form factors related to fluctuation fields}
	In contrast to the background form factor computations, the fluctuation approach (see \cite{Manrique:2009uh,Manrique:2010am,Christiansen:2012rx,Codello:2013fpa}  for early works) typically adopts a flat Euclidean background and organizes the effective average action in terms of $n$-point functions built from the fluctuation field,
	\begin{equation}\label{fluct}
	\Gamma_k = \sum_n \int \prod_{i=1}^n \frac{d^4p_i}{(2\pi)^4} \, \delta^4\left(\sum_i p_i \right) \, \Gamma_{\mu_1\nu_1 \cdots \mu_n\nu_n}^{(n)}(k;p_i) \, h_{\mu_1\nu_1}(p_1) \, \cdots \, h_{\mu_n\nu_n}(p_n) \, .  
	\end{equation}
	The $\Gamma^{(n)}$ depend on all momenta of the fluctuation fields and carry  non-trivial tensor structures. For instance, picking $\Gamma_k^{(2)}$ and focusing on the tensor structure associated with the transverse-traceless part of the graviton fluctuation allows to determine the momentum dependence of the graviton two-point function. Owed to the flat background, the evaluation of the trace can be done using standard momentum-space techniques. This technical advantage has allowed to resolve 
	much more information. Concretely, the full graviton two-point function as well as parts of the graviton three- and four-point functions have been investigated, corresponding to form factors with two, three and four powers of the curvature. Beyond that, several gravity-matter form factors have been calculated.
	
	The first work investigating the propagator of the graviton (and its attached Faddeev-Popov ghost) was \cite{Christiansen:2014raa}. In this work, the propagators were computed for the transverse-traceless component (transverse component for the ghost) only. A comparison with a derivative expansion showed that the latter is rather unstable at the lowest order, and one should either implement a constant anomalous dimension, or a bi-local approximation.
	
	The resolution of all components of the graviton and ghost propagator was achieved only recently \cite{Knorr:2021niv}. This revealed that the different graviton modes exhibit significant qualitative differences in their momentum dependence. On the other hand, the transverse and longitudinal mode of the ghost share a very similar propagator. Some general asymptotic relations of propagators have been derived, for example that for large momenta the two propagators of the ghost modes agree identically. Conceptually, \cite{Knorr:2021niv} elucidated the relation between momentum-dependent correlation functions and the form factors.
	
	The first investigation of the momentum dependence of the three-graviton vertex \cite{Christiansen:2015rva} also introduced a non-trivial condition, dubbed momentum locality, on the asymptotic scaling of the flow of correlation functions. Momentum locality is inherently related to a well-defined coarse graining. Some, but not all graviton correlation functions possess this property \cite{Christiansen:2015rva, Pawlowski:2020qer, Knorr:2021niv}. For example, the transverse-traceless part of the propagator as well as a specific projection of the three-graviton vertex display momentum locality, whereas the spin zero part of the propagator does not.
	
	The investigation of the four-graviton vertex \cite{Denz:2016qks} provided indications for an apparent convergence of the vertex expansion regarding the stability of the asymptotically safe fixed point. It also concluded that the generation of the operator $R_{\mu\nu}R^{\mu\nu}$ is suppressed compared with the generation of $R^2$.
	
	As a step beyond the form factor expansion about a flat background, \cite{Christiansen:2017bsy} resolved the momentum and (Ricci scalar) curvature dependence of the graviton propagator. Along the lines of the discussion in \autoref{sect.beyondflat}, this corresponds to a form factor expansion about a manifold with constant Ricci scalar which is necessary to compute exact scattering amplitudes on de Sitter and anti-de Sitter spaces \cite{Knorr:2020bjm, Ferrero:2021lhd}. Intriguingly, some of the couplings turn out to be approximately independent of the curvature.
	
	The most recent addition to the discussion of momentum dependence in quantum gravity is the computation of the graviton spectral function, both from Euclidean \cite{Bonanno:2021squ} and Lorentzian \cite{Fehre:2021eob} computations. This work represents a crucial step on the path to computing asymptotically safe scattering amplitudes, since all previous computations were carried out with Euclidean signature, whereas amplitudes probe genuine Lorentzian momentum configurations. This is particularly so for massless fields, where the on-shell condition $p^2=0$ in Euclidean signature implies a completely vanishing momentum vector. One key result of these investigations is that the (transverse-traceless) fluctuation graviton spectral function exists and is positive. A second key insight is that also non-perturbatively, there is only a single pole at vanishing mass, so that no additional ghosts appear in the spectrum.
	
	Coming to gravity-matter systems, an interesting relation for some gravitational diagrams contributing to the gluon propagator was found in \cite{Folkerts:2011jz}. More concisely, it was found that in the weak-gravity limit the gauge coupling does not receive contributions from the gravitational sector. A more detailed study of the two- and three-point functions of this system was put forward in \cite{Christiansen:2017cxa}. The scalar and fermionic propagators coupled to gravity were studied in \cite{Meibohm:2015twa}. In a particular gauge, there is no quantum correction to the propagator of a massless scalar. In \cite{Eichhorn:2018akn}, the concept of effective universality was introduced, and found to be present approximately at the asymptotically safe fixed point. The idea is that there are non-trivial relations related to diffeomorphism invariance that different vertices have to satisfy. In \cite{Draper:2020bop} it was remarked that such relations must be present for scattering amplitudes to possess a well-behaved high-energy limit. The momentum dependence of the graviton-fermion-fermion vertex was investigated in \cite{Eichhorn:2018ydy}, where indications were found that such a non-minimal coupling plays a sub-leading role. Finally, \cite{Burger:2019upn} extended the study of momentum- and curvature-dependent propagators to a gravity-scalar system.
	
	\subsection{Reconstruction from Monte Carlo data}
	\label{sec.63}
	An alternative avenue to obtain form factors from first principles is their reconstruction from Monte Carlo data. For the case of quantum gravity, this approach has been pioneered in \cite{Knorr:2018kog}, using correlation functions obtained within the Causal Dynamical Triangulation (CDT) program \cite{Ambjorn:2012jv,Loll:2019rdj} on a toroidal topology. The data used in the reconstruction utilized that the background geometry is flat in combination with the auto-correlation function of three-volume fluctuations,
	\begin{equation}\label{eq.autocorrel}
		\mathfrak V_2(t',t) = \langle \delta V_3(t') \, \delta V_3(t) \rangle \, \text{.}
	\end{equation}
		Here $t$ and $t'$ denote different times of the Cauchy slicing implemented in CDT.
	The latter has been measured for a specific set of bare couplings appearing in the path integral over causal geometries \cite{Ambjorn:2016fbd}. The idea of reconstructing the effective action from numerical data is clearly generally applicable to any quantum gravity approach that produces numerical data. So far, however, this algorithm has not been applied to other approaches besides CDT.
	
	Following the spirit of constructing the most general scattering amplitude compatible with the parameterized quantum effective action including form factors, one can identify all terms in $\Gamma$ that contribute to the correlator \eqref{eq.autocorrel}. Imposing diffeomorphism invariance, all relevant structures are contained in
	\begin{equation}\label{eq.g.ans}
	\Gamma \simeq \frac{1}{16 \pi \GN} \int d^4x \sqrt{g} \left[ 2 \Lambda - R - \frac{1}{6} R f_{RR}(\Box) R \right] \, .
	\end{equation}
	The flat background geometry fixes $\Lambda = 0$. Subsequently, one computes the auto-correlation function \eqref{eq.autocorrel} from \eqref{eq.g.ans}. 
	In \cite{Knorr:2018kog}, it was shown that this correlator reduces to
	\begin{equation}
	 \mathfrak V_2(t',t) \propto \sum_n{}^\prime \frac{1}{\lambda_n} \phi_n^\ast(t') \, \phi_n(t) \, ,
	\end{equation}
	where $\lambda_n$ are the eigenvalues and $\phi_n$ the eigenfunctions of the temporal part of the two-point function. The fitting procedure with the data from \cite{Ambjorn:2016fbd} then yielded the form factor
	\begin{equation}
	 f_{RR}(x) = -\frac{b^2}{x^2} \, ,
	\end{equation}
	for some real constant $b$. This result is notable since one generally does not expect such strong infrared non-localities, but rather a logarithm \cite{'tHooft:1974bx}. However, precisely this form has also been investigated in the context of cosmology to phenomenologically model dark energy \cite{Maggiore:2014sia, Belgacem:2017cqo}.

	\section{Conclusions}
	\label{sec.conclusion}
	The effective action $\Gamma$ provides a powerful tool to analyze quantum gravity effects within a broad range of quantum gravity programs building on the principles of quantum field theory. By construction, $\Gamma$ takes into account all quantum effects. Thus, quantum-corrected spacetimes may be obtained by solving the equations of motion provided by $\Gamma$ \cite{Becker:2014pea,Knorr:2022kqp}. Furthermore, scattering processes can be analyzed using tree-level Feynman diagrams built from the effective propagators and vertices derived from $\Gamma$. These properties turn the effective action in a pivotal element connecting fundamental computations to the resulting phenomenological consequences. 
	
	Within $\Gamma$, quantum effects manifest themselves in terms of form factors. These generalize the momentum-dependent couplings found in perturbative quantum field theory to arbitrary curved spacetimes in a background-independent way. Notably, they are capable of capturing genuine non-perturbative effects including, \eg{}, an anomalous dimension of the graviton propagator at trans-Planckian energy, $\eta_\infty = -2$.
	
	The form factor framework allows to determine the most general scattering amplitudes compatible with quantum field theory. The underlying two-step process is illustrated in \autoref{sec.ea} and \autoref{sec.scattering}. It first identifies all action monomials that contribute to a scattering process with a fixed set of external fields, and subsequently constructs the corresponding on-shell amplitude. Conceptually, working with these amplitudes is very appealing since they constitute physical observables that are gauge-independent, invariant with respect to field redefinitions, and depend on the essential couplings (in Weinberg's sense \cite{Weinberg:1980gg}) only, as discussed in \autoref{sec.tanh}. Moreover, their Lorentzian nature gives direct access to questions related to unitarity, causality, and the non-perturbative physics that could render gravity asymptotically safe. These features leave an imprint on the resulting low-energy effective field theory, and we exemplify some of the conjectured bounds in \autoref{sec.eft}.
		
	The close relation of $\Gamma$ to phenomenology motivates deriving this quantity (and in particular the form factors appearing in it) from first principles. Depending on the details of the microscopic theory, one may resort to standard perturbative quantization techniques. This applies, for instance, to the case of quadratic gravity discussed in \autoref{sect.52}. At the non-perturbative level, $\Gamma$ can be obtained by solving the Wetterich equation \cite{Wetterich:1992yh,Morris:1993qb} adapted to gravity \cite{Reuter:1996cp}. In this case, the effective action appears as the end point of a renormalization group trajectory in the limit where all quantum fluctuations have been integrated out. Conceptually, the Wetterich equation then determines the couplings and form factors in $\Gamma$ in terms of the relevant deformations of a renormalization group fixed point. The status of first principle computations seeking to determine the properties of form factors (foremost the graviton propagator in flat Euclidean backgrounds) along these lines is reviewed in \autoref{sec.hardcore}. Alternatively, one could start from correlation functions measured in Monte Carlo approaches to quantum gravity  to reverse-engineer the relevant terms in $\Gamma$ giving rise to these correlations. In the context of Causal Dynamical Triangulations \cite{Ambjorn:2012jv,Loll:2019rdj} this strategy has been pioneered in \cite{Knorr:2018kog}. As detailed in \autoref{sec.63}, this provided first clues that the form factors in the gravitational sector could contain inverse powers of the Laplacian.
	
	The present chapter illustrates that form factors play a crucial role in the gravitational Asymptotic Safety program. However, their relevance is, by no means, limited to it. In particular, they also constitute the key ingredients when formulating non-local, ghost-free gravity \cite{Biswas:2005qr,Biswas:2011ar,Buoninfante:2018xiw,Buoninfante:2020ctr} and perturbatively super-renormalizable theories of quantum gravity \cite{Modesto:2014lga,Modesto:2017hzl,Modesto:2017sdr}. This discussion is beyond the scope of this chapter and the interested reader is encouraged to consult the relevant chapters in this handbook for further information.

	\section*{Acknowledgements}
	Our understanding of form factors in the context of quantum gravity benefited from countless discussions with many colleagues. It is therefore a pleasure for us to thank 
	J.\ Ambj{\o}rn, 
	D.\ Becker, 
	M.\ Becker, 
	A.\ Bonanno,
	L.\ Bosma,
	L.\ Buoninfante, 
	J.\ Donoghue, 
	T.\ Draper,
	R.\ Ferrero, 
	G.\ Gubitosi,
	A.\ Koshelev, 
	S.\ Kumar, 
	P.\ Mannheim, 
	R.\ Ooijer,
	C.\ Pagani, 
	J.\ M.\ Pawlowski, 
	R.\ Percacci, 
	A.\ D.\ Pereira, 
	S.\ Pirlo,
	A.\ Platania, 
	M.\ Reichert, 
	M.\ Reuter, 
	M.\ Schiffer, 
	and C.\ Wetterich for sharing their thoughts with us. Moreover, we are in debt to A.\ G{\"o}rlich for sharing the CDT data underlying the analysis of \autoref{sec.63}. 
	B.\ K.\ acknowledges support by Perimeter Institute for Theoretical Physics and Nordita. Research at Perimeter Institute is supported in part by the Government of Canada through the Department of Innovation, Science and Economic Development and by the Province of Ontario through the Ministry of Colleges and Universities. Nordita is supported in part by NordForsk.
	
	\bibliographystyle{jhep}
	\bibliography{general_bib}

\providecommand{\href}[2]{#2}\begingroup\raggedright\begin{thebibliography}{100}

\bibitem{Percacci:2017fkn}
R.~Percacci, \emph{{An Introduction to Covariant Quantum Gravity and Asymptotic
  Safety}}, vol.~3 of \emph{100 Years of General Relativity}. World Scientific,
  2017, \href{https://doi.org/10.1142/10369}{10.1142/10369}.

\bibitem{Reuter:2019byg}
M.~Reuter and F.~Saueressig, \emph{{Quantum Gravity and the Functional
  Renormalization Group}}. Cambridge University Press, 2019.

\bibitem{Eichhorn:2018yfc}
A.~Eichhorn, \emph{{An asymptotically safe guide to quantum gravity and
  matter}}, \href{https://doi.org/10.3389/fspas.2018.00047}{\emph{Front.
  Astron. Space Sci.} {\bfseries 5} (2019) 47}
  [\href{https://arxiv.org/abs/1810.07615}{{\ttfamily 1810.07615}}].

\bibitem{Weinberg:1980gg}
S.~Weinberg, \emph{{Ultraviolet divergences in quantum theories of
  gravitation}}, {\emph{General Relativity: An Einstein centenary survey, Eds.
  Hawking, S.W., Israel, W; Cambridge University Press} (1979) 790}.

\bibitem{Reuter:1996cp}
M.~Reuter, \emph{{Nonperturbative evolution equation for quantum gravity}},
  \href{https://doi.org/10.1103/PhysRevD.57.971}{\emph{Phys.Rev.} {\bfseries
  D57} (1998) 971} [\href{https://arxiv.org/abs/hep-th/9605030}{{\ttfamily
  hep-th/9605030}}].

\bibitem{Wetterich:2019qzx}
C.~Wetterich, \emph{{Quantum scale symmetry}},
  \href{https://arxiv.org/abs/1901.04741}{{\ttfamily 1901.04741}}.

\bibitem{Knorr:2019atm}
B.~Knorr, C.~Ripken and F.~Saueressig, \emph{{Form Factors in Asymptotic
  Safety: conceptual ideas and computational toolbox}},
  \href{https://doi.org/10.1088/1361-6382/ab4a53}{\emph{Class. Quant. Grav.}
  {\bfseries 36} (2019) 234001}
  [\href{https://arxiv.org/abs/1907.02903}{{\ttfamily 1907.02903}}].

\bibitem{Camanho:2014apa}
X.~O. Camanho, J.~D. Edelstein, J.~Maldacena and A.~Zhiboedov, \emph{{Causality
  Constraints on Corrections to the Graviton Three-Point Coupling}},
  \href{https://doi.org/10.1007/JHEP02(2016)020}{\emph{JHEP} {\bfseries 02}
  (2016) 020} [\href{https://arxiv.org/abs/1407.5597}{{\ttfamily 1407.5597}}].

\bibitem{Cheung:2014ega}
C.~Cheung and G.~N. Remmen, \emph{{Infrared Consistency and the Weak Gravity
  Conjecture}}, \href{https://doi.org/10.1007/JHEP12(2014)087}{\emph{JHEP}
  {\bfseries 12} (2014) 087} [\href{https://arxiv.org/abs/1407.7865}{{\ttfamily
  1407.7865}}].

\bibitem{Tokuda:2020mlf}
J.~Tokuda, K.~Aoki and S.~Hirano, \emph{{Gravitational positivity bounds}},
  \href{https://doi.org/10.1007/JHEP11(2020)054}{\emph{JHEP} {\bfseries 11}
  (2020) 054} [\href{https://arxiv.org/abs/2007.15009}{{\ttfamily
  2007.15009}}].

\bibitem{Alberte:2020bdz}
L.~Alberte, C.~de~Rham, S.~Jaitly and A.~J. Tolley, \emph{{QED positivity
  bounds}}, \href{https://doi.org/10.1103/PhysRevD.103.125020}{\emph{Phys. Rev.
  D} {\bfseries 103} (2021) 125020}
  [\href{https://arxiv.org/abs/2012.05798}{{\ttfamily 2012.05798}}].

\bibitem{Alberte:2021dnj}
L.~Alberte, C.~de~Rham, S.~Jaitly and A.~J. Tolley, \emph{{Reverse
  Bootstrapping: IR Lessons for UV Physics}},
  \href{https://doi.org/10.1103/PhysRevLett.128.051602}{\emph{Phys. Rev. Lett.}
  {\bfseries 128} (2022) 051602}
  [\href{https://arxiv.org/abs/2111.09226}{{\ttfamily 2111.09226}}].

\bibitem{Wetterich:1992yh}
C.~Wetterich, \emph{{Exact evolution equation for the effective potential}},
  \href{https://doi.org/10.1016/0370-2693(93)90726-X}{\emph{Phys.Lett.}
  {\bfseries B301} (1993) 90}.

\bibitem{Morris:1993qb}
T.~R. Morris, \emph{{The Exact renormalization group and approximate
  solutions}}, \href{https://doi.org/10.1142/S0217751X94000972}{\emph{Int. J.
  Mod. Phys.} {\bfseries A9} (1994) 2411}
  [\href{https://arxiv.org/abs/hep-ph/9308265}{{\ttfamily hep-ph/9308265}}].

\bibitem{Reuter:1996ub}
M.~Reuter, \emph{{Effective average actions and nonperturbative evolution
  equations}},  in \emph{{5th Hellenic School and Workshops on Elementary
  Particle Physics}}, 2, 1996,
  \href{https://arxiv.org/abs/hep-th/9602012}{{\ttfamily hep-th/9602012}}.

\bibitem{Knorr:2018kog}
B.~Knorr and F.~Saueressig, \emph{{Towards reconstructing the quantum effective
  action of gravity}},
  \href{https://doi.org/10.1103/PhysRevLett.121.161304}{\emph{Phys. Rev. Lett.}
  {\bfseries 121} (2018) 161304}
  [\href{https://arxiv.org/abs/1804.03846}{{\ttfamily 1804.03846}}].

\bibitem{Donoghue:2017pgk}
J.~F. Donoghue, M.~M. Ivanov and A.~Shkerin, \emph{{EPFL Lectures on General
  Relativity as a Quantum Field Theory}},
  \href{https://arxiv.org/abs/1702.00319}{{\ttfamily 1702.00319}}.

\bibitem{Biswas:2005qr}
T.~Biswas, A.~Mazumdar and W.~Siegel, \emph{{Bouncing universes in
  string-inspired gravity}},
  \href{https://doi.org/10.1088/1475-7516/2006/03/009}{\emph{JCAP} {\bfseries
  0603} (2006) 009} [\href{https://arxiv.org/abs/hep-th/0508194}{{\ttfamily
  hep-th/0508194}}].

\bibitem{Biswas:2011ar}
T.~Biswas, E.~Gerwick, T.~Koivisto and A.~Mazumdar, \emph{{Towards singularity
  and ghost free theories of gravity}},
  \href{https://doi.org/10.1103/PhysRevLett.108.031101}{\emph{Phys. Rev. Lett.}
  {\bfseries 108} (2012) 031101}
  [\href{https://arxiv.org/abs/1110.5249}{{\ttfamily 1110.5249}}].

\bibitem{Buoninfante:2018xiw}
L.~Buoninfante, A.~S. Koshelev, G.~Lambiase and A.~Mazumdar, \emph{{Classical
  properties of non-local, ghost- and singularity-free gravity}},
  \href{https://doi.org/10.1088/1475-7516/2018/09/034}{\emph{JCAP} {\bfseries
  09} (2018) 034} [\href{https://arxiv.org/abs/1802.00399}{{\ttfamily
  1802.00399}}].

\bibitem{Buoninfante:2020ctr}
L.~Buoninfante, G.~Lambiase, Y.~Miyashita, W.~Takebe and M.~Yamaguchi,
  \emph{{Generalized ghost-free propagators in nonlocal field theories}},
  \href{https://doi.org/10.1103/PhysRevD.101.084019}{\emph{Phys. Rev. D}
  {\bfseries 101} (2020) 084019}
  [\href{https://arxiv.org/abs/2001.07830}{{\ttfamily 2001.07830}}].

\bibitem{Modesto:2014lga}
L.~Modesto and L.~Rachwal, \emph{{Super-renormalizable and finite gravitational
  theories}},
  \href{https://doi.org/10.1016/j.nuclphysb.2014.10.015}{\emph{Nucl. Phys. B}
  {\bfseries 889} (2014) 228}
  [\href{https://arxiv.org/abs/1407.8036}{{\ttfamily 1407.8036}}].

\bibitem{Modesto:2017hzl}
L.~Modesto, L.~Rachwa\l{} and I.~L. Shapiro, \emph{{Renormalization group in
  super-renormalizable quantum gravity}},
  \href{https://doi.org/10.1140/epjc/s10052-018-6035-2}{\emph{Eur. Phys. J. C}
  {\bfseries 78} (2018) 555}
  [\href{https://arxiv.org/abs/1704.03988}{{\ttfamily 1704.03988}}].

\bibitem{Modesto:2017sdr}
L.~Modesto and L.~Rachwal, \emph{{Nonlocal quantum gravity: A review}},
  \href{https://doi.org/10.1142/S0218271817300208}{\emph{Int. J. Mod. Phys.}
  {\bfseries D26} (2017) 1730020}.

\bibitem{Draper:2020bop}
T.~Draper, B.~Knorr, C.~Ripken and F.~Saueressig, \emph{{Finite Quantum Gravity
  Amplitudes: No Strings Attached}},
  \href{https://doi.org/10.1103/PhysRevLett.125.181301}{\emph{Phys. Rev. Lett.}
  {\bfseries 125} (2020) 181301}
  [\href{https://arxiv.org/abs/2007.00733}{{\ttfamily 2007.00733}}].

\bibitem{Draper:2020knh}
T.~Draper, B.~Knorr, C.~Ripken and F.~Saueressig, \emph{{Graviton-Mediated
  Scattering Amplitudes from the Quantum Effective Action}},
  \href{https://doi.org/10.1007/JHEP11(2020)136}{\emph{JHEP} {\bfseries 11}
  (2020) 136} [\href{https://arxiv.org/abs/2007.04396}{{\ttfamily
  2007.04396}}].

\bibitem{Knorr:2021iwv}
B.~Knorr, C.~Ripken and F.~Saueressig, \emph{{Form Factors in Quantum Gravity:
  Contrasting non-local, ghost-free gravity and Asymptotic Safety}},
  \href{https://doi.org/10.1393/ncc/i2022-22028-5}{\emph{Nuovo Cim. C}
  {\bfseries 45} (2022) 28} [\href{https://arxiv.org/abs/2111.12365}{{\ttfamily
  2111.12365}}].

\bibitem{Knorr:2022lzn}
B.~Knorr, S.~Pirlo, C.~Ripken and F.~Saueressig, \emph{{Cartographing
  gravity-mediated scattering amplitudes: scalars and photons}},
  \href{https://arxiv.org/abs/2205.01738}{{\ttfamily 2205.01738}}.

\bibitem{Codello:2012kq}
A.~Codello and O.~Zanusso, \emph{{On the non-local heat kernel expansion}},
  \href{https://doi.org/10.1063/1.4776234}{\emph{J. Math. Phys.} {\bfseries 54}
  (2013) 013513} [\href{https://arxiv.org/abs/1203.2034}{{\ttfamily
  1203.2034}}].

\bibitem{ParticleDataGroup:2016lqr}
{\scshape Particle Data Group} collaboration, \emph{{Review of Particle
  Physics}}, \href{https://doi.org/10.1088/1674-1137/40/10/100001}{\emph{Chin.
  Phys. C} {\bfseries 40} (2016) 100001}.

\bibitem{Bonanno:2020bil}
A.~Bonanno, A.~Eichhorn, H.~Gies, J.~M. Pawlowski, R.~Percacci, M.~Reuter
  et~al., \emph{{Critical reflections on asymptotically safe gravity}},
  \href{https://doi.org/10.3389/fphy.2020.00269}{\emph{Front. in Phys.}
  {\bfseries 8} (2020) 269} [\href{https://arxiv.org/abs/2004.06810}{{\ttfamily
  2004.06810}}].

\bibitem{Fulling:1992vm}
S.~A. Fulling, R.~C. King, B.~G. Wybourne and C.~J. Cummins, \emph{{Normal
  forms for tensor polynomials. 1: The Riemann tensor}},
  \href{https://doi.org/10.1088/0264-9381/9/5/003}{\emph{Class. Quant. Grav.}
  {\bfseries 9} (1992) 1151}.

\bibitem{Decanini:2008pr}
Y.~Decanini and A.~Folacci, \emph{{FKWC-bases and geometrical identities for
  classical and quantum field theories in curved spacetime}},
  \href{https://arxiv.org/abs/0805.1595}{{\ttfamily 0805.1595}}.

\bibitem{Chowdhury:2019kaq}
S.~D. Chowdhury, A.~Gadde, T.~Gopalka, I.~Halder, L.~Janagal and S.~Minwalla,
  \emph{{Classifying and constraining local four photon and four graviton
  S-matrices}}, \href{https://doi.org/10.1007/JHEP02(2020)114}{\emph{JHEP}
  {\bfseries 02} (2020) 114}
  [\href{https://arxiv.org/abs/1910.14392}{{\ttfamily 1910.14392}}].

\bibitem{Knorr:2020ckv}
B.~Knorr, \emph{{Lessons from conformally reduced quantum gravity}},
  \href{https://doi.org/10.1088/1361-6382/abd7c2}{\emph{Class. Quant. Grav.}
  {\bfseries 38} (2021) 065003}
  [\href{https://arxiv.org/abs/2010.00492}{{\ttfamily 2010.00492}}].

\bibitem{Barvinsky:1990up}
A.~Barvinsky and G.~Vilkovisky, \emph{{Covariant perturbation theory. 2: Second
  order in the curvature. General algorithms}},
  \href{https://doi.org/10.1016/0550-3213(90)90047-H}{\emph{Nucl. Phys. B}
  {\bfseries 333} (1990) 471}.

\bibitem{Barvinsky:1993en}
A.~O. Barvinsky, Y.~V. Gusev, V.~V. Zhytnikov and G.~A. Vilkovisky,
  \emph{{Covariant perturbation theory. 4. Third order in the curvature}},
  \href{https://arxiv.org/abs/0911.1168}{{\ttfamily 0911.1168}}.

\bibitem{Wegner_1974}
F.~J. Wegner, \emph{Some invariance properties of the renormalization group},
  \href{https://doi.org/10.1088/0022-3719/7/12/004}{\emph{Journal of Physics C:
  Solid State Physics} {\bfseries 7} (1974) 2098}.

\bibitem{Hawking:1979ig}
S.~W. Hawking and W.~E. Israel, \emph{{General Relativity - an Einstein
  Centenary Survey}}. Univ. Pr., Cambridge, UK, 1979.

\bibitem{Dietz:2013sba}
J.~A. Dietz and T.~R. Morris, \emph{{Redundant operators in the exact
  renormalisation group and in the f(R) approximation to asymptotic safety}},
  \href{https://doi.org/10.1007/JHEP07(2013)064}{\emph{JHEP} {\bfseries 07}
  (2013) 064} [\href{https://arxiv.org/abs/1306.1223}{{\ttfamily 1306.1223}}].

\bibitem{Baldazzi:2021ydj}
A.~Baldazzi, R.~B.~A. Zinati and K.~Falls, \emph{{Essential renormalisation
  group}}, \href{https://doi.org/10.21468/SciPostPhys.13.4.085}{\emph{SciPost
  Phys.} {\bfseries 13} (2022) 085}
  [\href{https://arxiv.org/abs/2105.11482}{{\ttfamily 2105.11482}}].

\bibitem{Baldazzi:2021orb}
A.~Baldazzi and K.~Falls, \emph{{Essential Quantum Einstein Gravity}},
  \href{https://doi.org/10.3390/universe7080294}{\emph{Universe} {\bfseries 7}
  (2021) 294} [\href{https://arxiv.org/abs/2107.00671}{{\ttfamily
  2107.00671}}].

\bibitem{'tHooft:1974bx}
G.~'t~Hooft and M.~J.~G. Veltman, \emph{{One loop divergencies in the theory of
  gravitation}}, {\emph{Annales Poincare Phys. Theor.} {\bfseries A20} (1974)
  69}.

\bibitem{Goroff:1985sz}
M.~H. Goroff and A.~Sagnotti, \emph{{QUANTUM GRAVITY AT TWO LOOPS}},
  \href{https://doi.org/10.1016/0370-2693(85)91470-4}{\emph{Phys. Lett.}
  {\bfseries B160} (1985) 81}.

\bibitem{Goroff:1985th}
M.~H. Goroff and A.~Sagnotti, \emph{{The Ultraviolet Behavior of Einstein
  Gravity}}, \href{https://doi.org/10.1016/0550-3213(86)90193-8}{\emph{Nucl.
  Phys.} {\bfseries B266} (1986) 709}.

\bibitem{Donoghue:1993eb}
J.~F. Donoghue, \emph{{Leading quantum correction to the Newtonian potential}},
  \href{https://doi.org/10.1103/PhysRevLett.72.2996}{\emph{Phys. Rev. Lett.}
  {\bfseries 72} (1994) 2996}
  [\href{https://arxiv.org/abs/gr-qc/9310024}{{\ttfamily gr-qc/9310024}}].

\bibitem{Donoghue:1994dn}
J.~F. Donoghue, \emph{{General relativity as an effective field theory: The
  leading quantum corrections}},
  \href{https://doi.org/10.1103/PhysRevD.50.3874}{\emph{Phys. Rev.} {\bfseries
  D50} (1994) 3874} [\href{https://arxiv.org/abs/gr-qc/9405057}{{\ttfamily
  gr-qc/9405057}}].

\bibitem{Bjerrum-Bohr:2014zsa}
N.~E.~J. Bjerrum-Bohr, J.~F. Donoghue, B.~R. Holstein, L.~Plant\'e and
  P.~Vanhove, \emph{{Bending of Light in Quantum Gravity}},
  \href{https://doi.org/10.1103/PhysRevLett.114.061301}{\emph{Phys. Rev. Lett.}
  {\bfseries 114} (2015) 061301}
  [\href{https://arxiv.org/abs/1410.7590}{{\ttfamily 1410.7590}}].

\bibitem{Bjerrum-Bohr:2017dxw}
N.~E.~J. Bjerrum-Bohr, B.~R. Holstein, J.~F. Donoghue, L.~Plant\'e and
  P.~Vanhove, \emph{{Illuminating Light Bending}},
  \href{https://doi.org/10.22323/1.292.0077}{\emph{PoS} {\bfseries CORFU2016}
  (2017) 077} [\href{https://arxiv.org/abs/1704.01624}{{\ttfamily
  1704.01624}}].

\bibitem{Polyakov:2007mm}
A.~M. Polyakov, \emph{{De Sitter space and eternity}},
  \href{https://doi.org/10.1016/j.nuclphysb.2008.01.002}{\emph{Nucl. Phys. B}
  {\bfseries 797} (2008) 199}
  [\href{https://arxiv.org/abs/0709.2899}{{\ttfamily 0709.2899}}].

\bibitem{Anderson:2013ila}
P.~R. Anderson and E.~Mottola, \emph{{Instability of global de Sitter space to
  particle creation}},
  \href{https://doi.org/10.1103/PhysRevD.89.104038}{\emph{Phys. Rev. D}
  {\bfseries 89} (2014) 104038}
  [\href{https://arxiv.org/abs/1310.0030}{{\ttfamily 1310.0030}}].

\bibitem{Anderson:2017hts}
P.~R. Anderson, E.~Mottola and D.~H. Sanders, \emph{{Decay of the de Sitter
  Vacuum}}, \href{https://doi.org/10.1103/PhysRevD.97.065016}{\emph{Phys. Rev.
  D} {\bfseries 97} (2018) 065016}
  [\href{https://arxiv.org/abs/1712.04522}{{\ttfamily 1712.04522}}].

\bibitem{Markkanen:2016aes}
T.~Markkanen and A.~Rajantie, \emph{{Massive scalar field evolution in de
  Sitter}}, \href{https://doi.org/10.1007/JHEP01(2017)133}{\emph{JHEP}
  {\bfseries 01} (2017) 133}
  [\href{https://arxiv.org/abs/1607.00334}{{\ttfamily 1607.00334}}].

\bibitem{Witten:2001kn}
E.~Witten, \emph{{Quantum gravity in de Sitter space}},  in \emph{{Strings
  2001: International Conference}}, 6, 2001,
  \href{https://arxiv.org/abs/hep-th/0106109}{{\ttfamily hep-th/0106109}}.

\bibitem{Bousso:2004tv}
R.~Bousso, \emph{{Cosmology and the S-matrix}},
  \href{https://doi.org/10.1103/PhysRevD.71.064024}{\emph{Phys. Rev. D}
  {\bfseries 71} (2005) 064024}
  [\href{https://arxiv.org/abs/hep-th/0412197}{{\ttfamily hep-th/0412197}}].

\bibitem{Marolf:2012kh}
D.~Marolf, I.~A. Morrison and M.~Srednicki, \emph{{Perturbative S-matrix for
  massive scalar fields in global de Sitter space}},
  \href{https://doi.org/10.1088/0264-9381/30/15/155023}{\emph{Class. Quant.
  Grav.} {\bfseries 30} (2013) 155023}
  [\href{https://arxiv.org/abs/1209.6039}{{\ttfamily 1209.6039}}].

\bibitem{Mandal:2019bdu}
S.~Mandal and S.~Banerjee, \emph{{Local description of S-matrix in quantum
  field theory in curved spacetime using Riemann-normal coordinate}},
  \href{https://doi.org/10.1140/epjp/s13360-021-02037-z}{\emph{Eur. Phys. J.
  Plus} {\bfseries 136} (2021) 1064}
  [\href{https://arxiv.org/abs/1908.06717}{{\ttfamily 1908.06717}}].

\bibitem{Giddings2013TheGS}
S.~B. Giddings, \emph{{The gravitational S-matrix: Erice lectures}},
  \href{https://doi.org/10.1142/9789814522489\_0005}{\emph{Subnucl. Ser.}
  {\bfseries 48} (2013) 93} [\href{https://arxiv.org/abs/1105.2036}{{\ttfamily
  1105.2036}}].

\bibitem{Knorr:2020bjm}
B.~Knorr and C.~Ripken, \emph{{Scattering amplitudes in affine gravity}},
  \href{https://doi.org/10.1103/PhysRevD.103.105019}{\emph{Phys. Rev. D}
  {\bfseries 103} (2021) 105019}
  [\href{https://arxiv.org/abs/2012.05144}{{\ttfamily 2012.05144}}].

\bibitem{Ferrero:2021lhd}
R.~Ferrero and C.~Ripken, \emph{{De Sitter scattering amplitudes in the Born
  approximation}},  \href{https://arxiv.org/abs/2112.03766}{{\ttfamily
  2112.03766}}.

\bibitem{Weinberg:1965nx}
S.~Weinberg, \emph{{Infrared photons and gravitons}},
  \href{https://doi.org/10.1103/PhysRev.140.B516}{\emph{Phys. Rev.} {\bfseries
  140} (1965) B516}.

\bibitem{Jin:1964zza}
Y.~S. Jin and A.~Martin, \emph{{Number of Subtractions in Fixed-Transfer
  Dispersion Relations}},
  \href{https://doi.org/10.1103/PhysRev.135.B1375}{\emph{Phys. Rev.} {\bfseries
  135} (1964) B1375}.

\bibitem{Stelle:1976gc}
K.~Stelle, \emph{{Renormalization of Higher Derivative Quantum Gravity}},
  \href{https://doi.org/10.1103/PhysRevD.16.953}{\emph{Phys. Rev. D} {\bfseries
  16} (1977) 953}.

\bibitem{Donoghue:2019ecz}
J.~F. Donoghue and G.~Menezes, \emph{{Arrow of Causality and Quantum Gravity}},
  \href{https://doi.org/10.1103/PhysRevLett.123.171601}{\emph{Phys. Rev. Lett.}
  {\bfseries 123} (2019) 171601}
  [\href{https://arxiv.org/abs/1908.04170}{{\ttfamily 1908.04170}}].

\bibitem{Anselmi:2018tmf}
D.~Anselmi and M.~Piva, \emph{{Quantum Gravity, Fakeons And Microcausality}},
  \href{https://doi.org/10.1007/JHEP11(2018)021}{\emph{JHEP} {\bfseries 11}
  (2018) 021} [\href{https://arxiv.org/abs/1806.03605}{{\ttfamily
  1806.03605}}].

\bibitem{Anselmi:2018bra}
D.~Anselmi, \emph{{Fakeons, Microcausality And The Classical Limit Of Quantum
  Gravity}}, \href{https://doi.org/10.1088/1361-6382/ab04c8}{\emph{Class.
  Quant. Grav.} {\bfseries 36} (2019) 065010}
  [\href{https://arxiv.org/abs/1809.05037}{{\ttfamily 1809.05037}}].

\bibitem{Donoghue:2021meq}
J.~F. Donoghue and G.~Menezes, \emph{{Causality and gravity}},
  \href{https://doi.org/10.1007/JHEP11(2021)010}{\emph{JHEP} {\bfseries 11}
  (2021) 010} [\href{https://arxiv.org/abs/2106.05912}{{\ttfamily
  2106.05912}}].

\bibitem{Donoghue:2018lmc}
J.~F. Donoghue and G.~Menezes, \emph{{Massive poles in Lee-Wick quantum field
  theory}}, \href{https://doi.org/10.1103/PhysRevD.99.065017}{\emph{Phys. Rev.
  D} {\bfseries 99} (2019) 065017}
  [\href{https://arxiv.org/abs/1812.03603}{{\ttfamily 1812.03603}}].

\bibitem{Reuter:2001ag}
M.~Reuter and F.~Saueressig, \emph{{Renormalization group flow of quantum
  gravity in the Einstein-Hilbert truncation}},
  \href{https://doi.org/10.1103/PhysRevD.65.065016}{\emph{Phys. Rev.}
  {\bfseries D65} (2002) 065016}
  [\href{https://arxiv.org/abs/hep-th/0110054}{{\ttfamily hep-th/0110054}}].

\bibitem{Platania:2022gtt}
A.~Platania, \emph{{Causality, unitarity and stability in quantum gravity: a
  non-perturbative perspective}},
  \href{https://doi.org/10.1007/JHEP09(2022)167}{\emph{JHEP} {\bfseries 09}
  (2022) 167} [\href{https://arxiv.org/abs/2206.04072}{{\ttfamily
  2206.04072}}].

\bibitem{Palti:2019pca}
E.~Palti, \emph{{The Swampland: Introduction and Review}},
  \href{https://doi.org/10.1002/prop.201900037}{\emph{Fortsch. Phys.}
  {\bfseries 67} (2019) 1900037}
  [\href{https://arxiv.org/abs/1903.06239}{{\ttfamily 1903.06239}}].

\bibitem{Basile:2021krr}
I.~Basile and A.~Platania, \emph{{Asymptotic Safety: Swampland or
  Wonderland?}}, \href{https://doi.org/10.3390/universe7100389}{\emph{Universe}
  {\bfseries 7} (2021) 389} [\href{https://arxiv.org/abs/2107.06897}{{\ttfamily
  2107.06897}}].

\bibitem{Becker:2017tcx}
D.~Becker, C.~Ripken and F.~Saueressig, \emph{{On avoiding Ostrogradski
  instabilities within Asymptotic Safety}},
  \href{https://doi.org/10.1007/JHEP12(2017)121}{\emph{JHEP} {\bfseries 12}
  (2017) 121} [\href{https://arxiv.org/abs/1709.09098}{{\ttfamily
  1709.09098}}].

\bibitem{Platania:2020knd}
A.~Platania and C.~Wetterich, \emph{{Non-perturbative unitarity and fictitious
  ghosts in quantum gravity}},
  \href{https://doi.org/10.1016/j.physletb.2020.135911}{\emph{Phys. Lett. B}
  {\bfseries 811} (2020) 135911}
  [\href{https://arxiv.org/abs/2009.06637}{{\ttfamily 2009.06637}}].

\bibitem{Pham:1985cr}
T.~N. Pham and T.~N. Truong, \emph{{Evaluation of the Derivative Quartic Terms
  of the Meson Chiral Lagrangian From Forward Dispersion Relation}},
  \href{https://doi.org/10.1103/PhysRevD.31.3027}{\emph{Phys. Rev. D}
  {\bfseries 31} (1985) 3027}.

\bibitem{Ananthanarayan:1994hf}
B.~Ananthanarayan, D.~Toublan and G.~Wanders, \emph{{Consistency of the chiral
  pion pion scattering amplitudes with axiomatic constraints}},
  \href{https://doi.org/10.1103/PhysRevD.51.1093}{\emph{Phys. Rev. D}
  {\bfseries 51} (1995) 1093}
  [\href{https://arxiv.org/abs/hep-ph/9410302}{{\ttfamily hep-ph/9410302}}].

\bibitem{Adams:2006sv}
A.~Adams, N.~Arkani-Hamed, S.~Dubovsky, A.~Nicolis and R.~Rattazzi,
  \emph{{Causality, analyticity and an IR obstruction to UV completion}},
  \href{https://doi.org/10.1088/1126-6708/2006/10/014}{\emph{JHEP} {\bfseries
  10} (2006) 014} [\href{https://arxiv.org/abs/hep-th/0602178}{{\ttfamily
  hep-th/0602178}}].

\bibitem{Harst:2011zx}
U.~Harst and M.~Reuter, \emph{{QED coupled to QEG}},
  \href{https://doi.org/10.1007/JHEP05(2011)119}{\emph{JHEP} {\bfseries 1105}
  (2011) 119} [\href{https://arxiv.org/abs/1101.6007}{{\ttfamily 1101.6007}}].

\bibitem{deBrito:2020dta}
G.~P. de~Brito, A.~Eichhorn and M.~Schiffer, \emph{{Light charged fermions in
  quantum gravity}},
  \href{https://doi.org/10.1016/j.physletb.2021.136128}{\emph{Phys. Lett. B}
  {\bfseries 815} (2021) 136128}
  [\href{https://arxiv.org/abs/2010.00605}{{\ttfamily 2010.00605}}].

\bibitem{Gies:2020xuh}
H.~Gies and J.~Ziebell, \emph{{Asymptotically Safe QED}},
  \href{https://doi.org/10.1140/epjc/s10052-020-8171-8}{\emph{Eur. Phys. J. C}
  {\bfseries 80} (2020) 607}
  [\href{https://arxiv.org/abs/2005.07586}{{\ttfamily 2005.07586}}].

\bibitem{Arkani-Hamed:2006emk}
N.~Arkani-Hamed, L.~Motl, A.~Nicolis and C.~Vafa, \emph{{The String landscape,
  black holes and gravity as the weakest force}},
  \href{https://doi.org/10.1088/1126-6708/2007/06/060}{\emph{JHEP} {\bfseries
  06} (2007) 060} [\href{https://arxiv.org/abs/hep-th/0601001}{{\ttfamily
  hep-th/0601001}}].

\bibitem{Drummond:1979pp}
I.~T. Drummond and S.~J. Hathrell, \emph{{QED Vacuum Polarization in a
  Background Gravitational Field and Its Effect on the Velocity of Photons}},
  \href{https://doi.org/10.1103/PhysRevD.22.343}{\emph{Phys. Rev. D} {\bfseries
  22} (1980) 343}.

\bibitem{Eichhorn:2012va}
A.~Eichhorn, \emph{{Quantum-gravity-induced matter self-interactions in the
  asymptotic-safety scenario}},
  \href{https://doi.org/10.1103/PhysRevD.86.105021}{\emph{Phys. Rev.}
  {\bfseries D86} (2012) 105021}
  [\href{https://arxiv.org/abs/1204.0965}{{\ttfamily 1204.0965}}].

\bibitem{Laporte:2021kyp}
C.~Laporte, A.~D. Pereira, F.~Saueressig and J.~Wang, \emph{{Scalar-tensor
  theories within Asymptotic Safety}},
  \href{https://doi.org/10.1007/JHEP12(2021)001}{\emph{JHEP} {\bfseries 12}
  (2021) 001} [\href{https://arxiv.org/abs/2110.09566}{{\ttfamily
  2110.09566}}].

\bibitem{Becker:2014pea}
D.~Becker and M.~Reuter, \emph{{Towards a $C$-function in 4D quantum gravity}},
  \href{https://doi.org/10.1007/JHEP03(2015)065}{\emph{JHEP} {\bfseries 03}
  (2015) 065} [\href{https://arxiv.org/abs/1412.0468}{{\ttfamily 1412.0468}}].

\bibitem{Edelstein:2021jyu}
J.~D. Edelstein, R.~Ghosh, A.~Laddha and S.~Sarkar, \emph{{Causality
  constraints in Quadratic Gravity}},
  \href{https://doi.org/10.1007/JHEP09(2021)150}{\emph{JHEP} {\bfseries 09}
  (2021) 150} [\href{https://arxiv.org/abs/2107.07424}{{\ttfamily
  2107.07424}}].

\bibitem{Bern:2021ppb}
Z.~Bern, D.~Kosmopoulos and A.~Zhiboedov, \emph{{Gravitational effective field
  theory islands, low-spin dominance, and the four-graviton amplitude}},
  \href{https://doi.org/10.1088/1751-8121/ac0e51}{\emph{J. Phys. A} {\bfseries
  54} (2021) 344002} [\href{https://arxiv.org/abs/2103.12728}{{\ttfamily
  2103.12728}}].

\bibitem{Niedermaier:2006wt}
M.~Niedermaier and M.~Reuter, \emph{{The Asymptotic Safety Scenario in Quantum
  Gravity}}, \href{https://doi.org/10.12942/lrr-2006-5}{\emph{Living Rev.Rel.}
  {\bfseries 9} (2006) 5}.

\bibitem{Codello:2008vh}
A.~Codello, R.~Percacci and C.~Rahmede, \emph{{Investigating the Ultraviolet
  Properties of Gravity with a Wilsonian Renormalization Group Equation}},
  \href{https://doi.org/10.1016/j.aop.2008.08.008}{\emph{Annals Phys.}
  {\bfseries 324} (2009) 414}
  [\href{https://arxiv.org/abs/0805.2909}{{\ttfamily 0805.2909}}].

\bibitem{Reuter:2012id}
M.~Reuter and F.~Saueressig, \emph{{Quantum Einstein Gravity}},
  \href{https://doi.org/10.1088/1367-2630/14/5/055022}{\emph{New J.Phys.}
  {\bfseries 14} (2012) 055022}
  [\href{https://arxiv.org/abs/1202.2274}{{\ttfamily 1202.2274}}].

\bibitem{Dupuis:2020fhh}
N.~Dupuis, L.~Canet, A.~Eichhorn, W.~Metzner, J.~M. Pawlowski, M.~Tissier
  et~al., \emph{{The nonperturbative functional renormalization group and its
  applications}},
  \href{https://doi.org/10.1016/j.physrep.2021.01.001}{\emph{Physics Reports}
  (2020) } [\href{https://arxiv.org/abs/2006.04853}{{\ttfamily 2006.04853}}].

\bibitem{me1}
F.~Saueressig, \emph{{The Functional Renormalization Group in Quantum
  Gravity}}, .

\bibitem{Pawlowski:2020qer}
J.~M. Pawlowski and M.~Reichert, \emph{{Quantum Gravity: A Fluctuating Point of
  View}}, \href{https://doi.org/10.3389/fphy.2020.551848}{\emph{Front. in
  Phys.} {\bfseries 8} (2021) 551848}
  [\href{https://arxiv.org/abs/2007.10353}{{\ttfamily 2007.10353}}].

\bibitem{Decanini:2005gt}
Y.~Decanini and A.~Folacci, \emph{{Off-diagonal coefficients of the
  Dewitt-Schwinger and Hadamard representations of the Feynman propagator}},
  \href{https://doi.org/10.1103/PhysRevD.73.044027}{\emph{Phys. Rev.}
  {\bfseries D73} (2006) 044027}
  [\href{https://arxiv.org/abs/gr-qc/0511115}{{\ttfamily gr-qc/0511115}}].

\bibitem{Benedetti:2010nr}
D.~Benedetti, K.~Groh, P.~F. Machado and F.~Saueressig, \emph{{The Universal RG
  Machine}}, \href{https://doi.org/10.1007/JHEP06(2011)079}{\emph{JHEP}
  {\bfseries 1106} (2011) 079}
  [\href{https://arxiv.org/abs/1012.3081}{{\ttfamily 1012.3081}}].

\bibitem{Codello:2010mj}
A.~Codello, \emph{{Polyakov Effective Action from Functional Renormalization
  Group Equation}},
  \href{https://doi.org/10.1016/j.aop.2010.04.013}{\emph{Annals Phys.}
  {\bfseries 325} (2010) 1727}
  [\href{https://arxiv.org/abs/1004.2171}{{\ttfamily 1004.2171}}].

\bibitem{Satz:2010uu}
A.~Satz, A.~Codello and F.~Mazzitelli, \emph{{Low energy Quantum Gravity from
  the Effective Average Action}},
  \href{https://doi.org/10.1103/PhysRevD.82.084011}{\emph{Phys. Rev. D}
  {\bfseries 82} (2010) 084011}
  [\href{https://arxiv.org/abs/1006.3808}{{\ttfamily 1006.3808}}].

\bibitem{Codello:2015oqa}
A.~Codello, R.~Percacci, L.~Rachwa\l{} and A.~Tonero, \emph{{Computing the
  Effective Action with the Functional Renormalization Group}},
  \href{https://doi.org/10.1140/epjc/s10052-016-4063-3}{\emph{Eur. Phys. J. C}
  {\bfseries 76} (2016) 226}
  [\href{https://arxiv.org/abs/1505.03119}{{\ttfamily 1505.03119}}].

\bibitem{Codello:2015mba}
A.~Codello and R.~K. Jain, \emph{{On the covariant formalism of the effective
  field theory of gravity and leading order corrections}},
  \href{https://doi.org/10.1088/0264-9381/33/22/225006}{\emph{Class. Quant.
  Grav.} {\bfseries 33} (2016) 225006}
  [\href{https://arxiv.org/abs/1507.06308}{{\ttfamily 1507.06308}}].

\bibitem{Ohta:2020bsc}
N.~Ohta and L.~Rachwal, \emph{{Effective action from the functional
  renormalization group}},
  \href{https://doi.org/10.1140/epjc/s10052-020-8325-8}{\emph{Eur. Phys. J. C}
  {\bfseries 80} (2020) 877}
  [\href{https://arxiv.org/abs/2002.10839}{{\ttfamily 2002.10839}}].

\bibitem{Knorr:2021niv}
B.~Knorr and M.~Schiffer, \emph{{Non-Perturbative Propagators in Quantum
  Gravity}}, \href{https://doi.org/10.3390/universe7070216}{\emph{Universe}
  {\bfseries 7} (2021) 216} [\href{https://arxiv.org/abs/2105.04566}{{\ttfamily
  2105.04566}}].

\bibitem{Codello:2015pga}
A.~Codello and R.~K. Jain, \emph{{On the covariant formalism of the effective
  field theory of gravity and its cosmological implications}},
  \href{https://doi.org/10.1088/1361-6382/aa549d}{\emph{Class. Quant. Grav.}
  {\bfseries 34} (2017) 035015}
  [\href{https://arxiv.org/abs/1507.07829}{{\ttfamily 1507.07829}}].

\bibitem{Bosma:2019aiu}
L.~Bosma, B.~Knorr and F.~Saueressig, \emph{{Resolving Spacetime Singularities
  within Asymptotic Safety}},
  \href{https://doi.org/10.1103/PhysRevLett.123.101301}{\emph{Phys. Rev. Lett.}
  {\bfseries 123} (2019) 101301}
  [\href{https://arxiv.org/abs/1904.04845}{{\ttfamily 1904.04845}}].

\bibitem{Bonanno:2000ep}
A.~Bonanno and M.~Reuter, \emph{{Renormalization group improved black hole
  space-times}}, \href{https://doi.org/10.1103/PhysRevD.62.043008}{\emph{Phys.
  Rev.} {\bfseries D62} (2000) 043008}
  [\href{https://arxiv.org/abs/hep-th/0002196}{{\ttfamily hep-th/0002196}}].

\bibitem{Meibohm:2015twa}
J.~Meibohm, J.~M. Pawlowski and M.~Reichert, \emph{{Asymptotic safety of
  gravity-matter systems}},
  \href{https://doi.org/10.1103/PhysRevD.93.084035}{\emph{Phys. Rev.}
  {\bfseries D93} (2016) 084035}
  [\href{https://arxiv.org/abs/1510.07018}{{\ttfamily 1510.07018}}].

\bibitem{Manrique:2009uh}
E.~Manrique and M.~Reuter, \emph{{Bimetric Truncations for Quantum Einstein
  Gravity and Asymptotic Safety}},
  \href{https://doi.org/10.1016/j.aop.2009.11.009}{\emph{Annals Phys.}
  {\bfseries 325} (2010) 785}
  [\href{https://arxiv.org/abs/0907.2617}{{\ttfamily 0907.2617}}].

\bibitem{Manrique:2010am}
E.~Manrique, M.~Reuter and F.~Saueressig, \emph{{Bimetric Renormalization Group
  Flows in Quantum Einstein Gravity}},
  \href{https://doi.org/10.1016/j.aop.2010.11.006}{\emph{Annals Phys.}
  {\bfseries 326} (2011) 463}
  [\href{https://arxiv.org/abs/1006.0099}{{\ttfamily 1006.0099}}].

\bibitem{Christiansen:2012rx}
N.~Christiansen, D.~F. Litim, J.~M. Pawlowski and A.~Rodigast, \emph{{Fixed
  points and infrared completion of quantum gravity}},
  \href{https://doi.org/10.1016/j.physletb.2013.11.025}{\emph{Phys.Lett.}
  {\bfseries B728} (2014) 114}
  [\href{https://arxiv.org/abs/1209.4038}{{\ttfamily 1209.4038}}].

\bibitem{Codello:2013fpa}
A.~Codello, G.~D'Odorico and C.~Pagani, \emph{{Consistent closure of
  renormalization group flow equations in quantum gravity}},
  \href{https://doi.org/10.1103/PhysRevD.89.081701}{\emph{Phys.Rev.} {\bfseries
  D89} (2014) 081701} [\href{https://arxiv.org/abs/1304.4777}{{\ttfamily
  1304.4777}}].

\bibitem{Christiansen:2014raa}
N.~Christiansen, B.~Knorr, J.~M. Pawlowski and A.~Rodigast, \emph{{Global Flows
  in Quantum Gravity}},
  \href{https://doi.org/10.1103/PhysRevD.93.044036}{\emph{Phys. Rev.}
  {\bfseries D93} (2016) 044036}
  [\href{https://arxiv.org/abs/1403.1232}{{\ttfamily 1403.1232}}].

\bibitem{Christiansen:2015rva}
N.~Christiansen, B.~Knorr, J.~Meibohm, J.~M. Pawlowski and M.~Reichert,
  \emph{{Local Quantum Gravity}},
  \href{https://doi.org/10.1103/PhysRevD.92.121501}{\emph{Phys. Rev.}
  {\bfseries D92} (2015) 121501}
  [\href{https://arxiv.org/abs/1506.07016}{{\ttfamily 1506.07016}}].

\bibitem{Denz:2016qks}
T.~Denz, J.~M. Pawlowski and M.~Reichert, \emph{{Towards apparent convergence
  in asymptotically safe quantum gravity}},
  \href{https://doi.org/10.1140/epjc/s10052-018-5806-0}{\emph{Eur. Phys. J.}
  {\bfseries C78} (2018) 336}
  [\href{https://arxiv.org/abs/1612.07315}{{\ttfamily 1612.07315}}].

\bibitem{Christiansen:2017bsy}
N.~Christiansen, K.~Falls, J.~M. Pawlowski and M.~Reichert, \emph{{Curvature
  dependence of quantum gravity}},
  \href{https://doi.org/10.1103/PhysRevD.97.046007}{\emph{Phys. Rev.}
  {\bfseries D97} (2018) 046007}
  [\href{https://arxiv.org/abs/1711.09259}{{\ttfamily 1711.09259}}].

\bibitem{Bonanno:2021squ}
A.~Bonanno, T.~Denz, J.~M. Pawlowski and M.~Reichert, \emph{{Reconstructing the
  graviton}},
  \href{https://doi.org/10.21468/SciPostPhys.12.1.001}{\emph{SciPost Phys.}
  {\bfseries 12} (2022) 001}
  [\href{https://arxiv.org/abs/2102.02217}{{\ttfamily 2102.02217}}].

\bibitem{Fehre:2021eob}
J.~Fehre, D.~F. Litim, J.~M. Pawlowski and M.~Reichert, \emph{{Lorentzian
  quantum gravity and the graviton spectral function}},
  \href{https://arxiv.org/abs/2111.13232}{{\ttfamily 2111.13232}}.

\bibitem{Folkerts:2011jz}
S.~Folkerts, D.~F. Litim and J.~M. Pawlowski, \emph{{Asymptotic freedom of
  Yang-Mills theory with gravity}},
  \href{https://doi.org/10.1016/j.physletb.2012.02.002}{\emph{Phys.Lett.}
  {\bfseries B709} (2012) 234}
  [\href{https://arxiv.org/abs/1101.5552}{{\ttfamily 1101.5552}}].

\bibitem{Christiansen:2017cxa}
N.~Christiansen, D.~F. Litim, J.~M. Pawlowski and M.~Reichert,
  \emph{{Asymptotic safety of gravity with matter}},
  \href{https://doi.org/10.1103/PhysRevD.97.106012}{\emph{Phys. Rev.}
  {\bfseries D97} (2018) 106012}
  [\href{https://arxiv.org/abs/1710.04669}{{\ttfamily 1710.04669}}].

\bibitem{Eichhorn:2018akn}
A.~Eichhorn, P.~Labus, J.~M. Pawlowski and M.~Reichert, \emph{{Effective
  universality in quantum gravity}},
  \href{https://doi.org/10.21468/SciPostPhys.5.4.031}{\emph{SciPost Phys.}
  {\bfseries 5} (2018) 031} [\href{https://arxiv.org/abs/1804.00012}{{\ttfamily
  1804.00012}}].

\bibitem{Eichhorn:2018ydy}
A.~Eichhorn, S.~Lippoldt, J.~M. Pawlowski, M.~Reichert and M.~Schiffer,
  \emph{{How perturbative is quantum gravity?}},
  \href{https://doi.org/10.1016/j.physletb.2019.01.071}{\emph{Phys. Lett.}
  {\bfseries B792} (2019) 310}
  [\href{https://arxiv.org/abs/1810.02828}{{\ttfamily 1810.02828}}].

\bibitem{Burger:2019upn}
B.~Bürger, J.~M. Pawlowski, M.~Reichert and B.-J. Schaefer, \emph{{Curvature
  dependence of quantum gravity with scalars}},
  \href{https://arxiv.org/abs/1912.01624}{{\ttfamily 1912.01624}}.

\bibitem{Ambjorn:2012jv}
J.~Ambjørn, A.~Görlich, J.~Jurkiewicz and R.~Loll, \emph{{Nonperturbative
  Quantum Gravity}},
  \href{https://doi.org/10.1016/j.physrep.2012.03.007}{\emph{Phys. Rept.}
  {\bfseries 519} (2012) 127}
  [\href{https://arxiv.org/abs/1203.3591}{{\ttfamily 1203.3591}}].

\bibitem{Loll:2019rdj}
R.~Loll, \emph{{Quantum Gravity from Causal Dynamical Triangulations: A
  Review}}, \href{https://doi.org/10.1088/1361-6382/ab57c7}{\emph{Class. Quant.
  Grav.} {\bfseries 37} (2020) 013002}
  [\href{https://arxiv.org/abs/1905.08669}{{\ttfamily 1905.08669}}].

\bibitem{Ambjorn:2016fbd}
J.~Ambjørn, Z.~Drogosz, J.~Gizbert-Studnicki, A.~Görlich, J.~Jurkiewicz and
  D.~Nemeth, \emph{{Impact of topology in causal dynamical triangulations
  quantum gravity}},
  \href{https://doi.org/10.1103/PhysRevD.94.044010}{\emph{Phys. Rev.}
  {\bfseries D94} (2016) 044010}
  [\href{https://arxiv.org/abs/1604.08786}{{\ttfamily 1604.08786}}].

\bibitem{Maggiore:2014sia}
M.~Maggiore and M.~Mancarella, \emph{{Nonlocal gravity and dark energy}},
  \href{https://doi.org/10.1103/PhysRevD.90.023005}{\emph{Phys. Rev.}
  {\bfseries D90} (2014) 023005}
  [\href{https://arxiv.org/abs/1402.0448}{{\ttfamily 1402.0448}}].

\bibitem{Belgacem:2017cqo}
E.~Belgacem, Y.~Dirian, S.~Foffa and M.~Maggiore, \emph{{Nonlocal gravity.
  Conceptual aspects and cosmological predictions}},
  \href{https://doi.org/10.1088/1475-7516/2018/03/002}{\emph{JCAP} {\bfseries
  1803} (2018) 002} [\href{https://arxiv.org/abs/1712.07066}{{\ttfamily
  1712.07066}}].

\bibitem{Knorr:2022kqp}
B.~Knorr and A.~Platania, \emph{{Sifting quantum black holes through the
  principle of least action}},
  \href{https://doi.org/10.1103/PhysRevD.106.L021901}{\emph{Phys. Rev. D}
  {\bfseries 106} (2022) L021901}
  [\href{https://arxiv.org/abs/2202.01216}{{\ttfamily 2202.01216}}].

\end{thebibliography}\endgroup
	
\end{document}